\documentclass[structabstract]{aa}  
\usepackage{aalongtable,lscape}
\usepackage{graphicx}
\usepackage{txfonts}
\begin{document}

\title{CARMENES input catalogue of M dwarfs}  
\subtitle{I. Low-resolution spectroscopy with CAFOS}
\titlerunning{Low-resolution spectroscopy with CAFOS}
\author{F.~J. Alonso-Floriano\inst{1}
        \and
        J.~C. Morales\inst{2,3}
        \and
        J.~A. Caballero\inst{4}
        \and
        D. Montes\inst{1}
        \and
        A. Klutsch\inst{5,1} 
        \and 
        R. Mundt\inst{6} 
        \and 
        \\ M. Cort\'es-Contreras\inst{1}
        \and
        I. Ribas\inst{2}
        \and
        A. Reiners\inst{7}
        \and
        P.~J. Amado\inst{8}
        \and
        A. Quirrenbach\inst{9}
        \and
        S.~V. Jeffers\inst{7}
        }
\institute{
        Departamento de Astrof\'{\i}sica y Ciencias de la Atm\'osfera, Facultad de Ciencias F\'{\i}sicas, Universidad Complutense de Madrid, 28040 Madrid, Spain, \email{fjalonso@fis.ucm.es}
        \and
         Institut de Ci\`encies de l'Espai (CSIC-IEEC), Campus UAB, Facultat Ci\`encies, Torre C5\,-\,parell\,-\,2$^a$, 08193 Bellaterra, Barcelona, Spain
        \and
        LESIA-Observatoire de Paris, CNRS, UPMC Univ. Paris 06, Univ. Paris-Diderot, 5 Pl. Jules Janssen, 92195 Meudon Cedex, France
        \and
        Departamento de Astrof\'isica, Centro de Astrobiolog\'ia (CSIC--INTA), PO~Box~78, 28691 Villanueva de la Ca\~nada, Madrid, Spain
        \and
        INAF - Osservatorio Astrofisico di Catania, via S. Sofia 78, 95123 Catania, Italy
        \and
        Max-Planck-Institut f\"ur Astronomie, K\"onigstuhl 17, 69117 Heidelberg, Germany
        \and
        Institut f\"ur Astrophysik, Friedrich-Hund-Platz 1, 37077 G\"ottingen, Germany
        \and
        Instituto de Astrof\'isica de Andaluc\'ia (CSIC), Glorieta de la Astronom\'ia s/n, 18008 Granada, Spain 
        \and
        Landessternwarte, Zentrum f\"ur Astronomie der Universit\"at Heidelberg, K\"onigstuhl 12, 69117 Heidelberg, Germany 
        }
\date{Received 4 February 2015; accepted 18 February 2015}

\abstract
{CARMENES is a stabilised, high-resolution, double-channel spectrograph at the 3.5\,m Calar Alto telescope. It is optimally designed for radial-velocity surveys of M dwarfs with potentially habitable Earth-mass planets.} 
{We prepare a list of the brightest, single M dwarfs in each spectral subtype observable from the northern hemisphere, from which we will select the best planet-hunting targets for CARMENES.} 
{In this first paper on the preparation of our input catalogue, we compiled a large amount of public data and collected low-resolution optical spectroscopy with CAFOS at the 2.2\,m Calar Alto telescope for 753 stars.
We derived accurate spectral types using a dense grid of standard stars, a double least-squares minimisation technique, and 31 spectral indices previously defined by other authors.
Additionally, we quantified surface gravity, metallicity, and chromospheric activity for all the stars in our sample.}
{We calculated spectral types for all 753 stars, of which {305} are new and {448} are revised.
We measured pseudo-equivalent widths of H$\alpha$ for all the stars in our sample, concluded that chromospheric activity does not affect spectral typing from our indices, and tabulated {49} stars that had been reported to be young stars in open clusters, moving groups, and stellar associations.
Of the 753 stars, two are new subdwarf candidates, three are T~Tauri stars, 25 are giants, 44 are K dwarfs, and 679 are M dwarfs.
Many of the 261 investigated dwarfs in the range M4.0--8.0\,V are among the brightest stars known in their spectral subtype.
}
{This collection of low-resolution spectroscopic data serves as a candidate target list for the CARMENES survey and can be highly valuable for other radial-velocity surveys of M dwarfs and for studies of cool dwarfs in the solar neighbourhood.}
\maketitle
%

\section{Introduction}
\label{section.introduction}

The {\em Calar Alto high-Resolution search for M dwarfs with Exo-earths with Near-infrared and optical \'Echelle Spectrographs} (hereafter CARMENES\footnote{{\tt http://carmenes.caha.es} -- Pronunciation: {\tt /k$\alpha$r'$\cdot$m$\varepsilon$n$\cdot\varepsilon$s/}}) is a next-generation instrument close to completion for the Zeiss 3.5\,m Calar Alto telescope, which is located in the Sierra de Los Filabres, Almer\'{\i}a, in southern Spain, at a height of about 2200\,m (S\'anchez et~al. 2007, 2008).
CARMENES is the name of used for the instrument, the consortium of 11 German and Spanish institutions that builds it, and of the scientific project to be carried out during guaranteed time observations. 
The instrument consists of two separated, highly stable, fibre-fed spectrographs covering the wavelength ranges from 0.55 to 0.95\,$\mu$m and from 0.95 to 1.70\,$\mu$m at spectral resolution R $\approx$ 82\,000, each of which shall perform high-accuracy radial-velocity measurements with long-term stability of $\sim$1\,m\,s$^{-1}$ (Quirrenbach et~al. 2010, 2012, 2014, and references therein; Amado et~al. 2013). 
First light is scheduled for the summer of 2015, followed by the commission of the instrument in the second half of that year.

The main scientific objective for CARMENES is the search for very low-mass planets (i.e., super- and exo-earths) orbiting mid- to late-M dwarfs, including a sample of moderately active M-dwarf stars.
Dwarf stars of M spectral type have effective temperatures between 2300 and 3900\,K (Kirkpatrick et~al. 2005; Rajpurohit et~al. 2013).
For stars with ages greater than that of the Hyades, of about 0.6\,Ga, these effective temperatures translate in the main sequence into a mass interval from 0.09 to 0.55\,$M_\odot$, approximately (Baraffe et~al. 1998; Chabrier et~al. 2000; Allard et~al. 2011). 
Of particular interest is the detection of very low-mass planets in the stellar habitable zone, the region around the star within which a planet can support liquid water (Kasting et~al. 1993; Joshi et~al. 1997; Lammer et~al. 2007; Tarter et~al. 2007; Scalo et~al. 2007).
In principle, the lower the mass of a host star, the higher the radial-velocity amplitude velocity induced (i.e., $K_{\rm star}$ is proportional to $M_{\rm planet} ~ a^{-1/2} ~ (M_{\rm star} + M_{\rm planet})^{-1/2} \approx ({a ~ M_{\rm star}})^{-1/2}$ when $M_{\rm star} \gg M_{\rm planet}$).
In addition, the lower luminosity of an M dwarf with respect to a star of earlier spectral type causes its habitable zone to be located very close to the host star,
which makes detecting habitable planets around M dwarfs (at $\sim$0.1\,au) easier than detections around solar-like stars (at $\sim$1\,au).  

From transit surveys with the NASA 0.95\,m {\em Kepler} space observatory, very small planet candidates are found to be relatively more abundant than Jupiter-type candidates as the host stellar mass decreases (Howard et~al. 2012; Dressing \& Charbonneau 2013, 2015; Kopparapu 2013; Kopparapu et~al. 2013). 
For early-M dwarf stars in the field, some radial-velocity studies have already been carried out (ESO CES, UVES and HARPS by Zechmeister et~al. 2009, 2013; CRIRES by Bean et~al. 2010; HARPS by Bonfils et~al. 2013), but the much-sought value of $\eta_\oplus$, that
is, the relative abundance of Earth-type planets in the habitable zone, is as yet only poorly constrained from radial-velocity data (e.g., $\eta_\oplus$ = 0.41$^{+0.54}_{-0.13}$ from Bonfils et~al. 2013).

Highly stable, high-resolution spectrographs in the near-infrared currently under construction, such as SPIRou (Artigau et~al. 2014), IRD (Kotani et~al. 2014), HPF (Mahadevan et~al. 2014), and CARMENES, are preferable over visible for targets with spectral types M4\,V or later (see Table~1 in Crossfield 2014). 
This is because the spectral energy distribution of late-M dwarfs approximately peaks at 1.0--1.2\,$\mu$m (Reiners et~al. 2010), while HARPS and its copy in the northern hemisphere, HARPS-N, cover the wavelength interval from 0.38 to 0.69\,$\mu$m.  
That faintness in the optical is quantitatively illustrated with the tabulated $V$ magnitudes of the brightest M dwarfs in the northern hemisphere (\object{HD~79210}/GJ~338\,A, \object{HD~79211}/GJ~338\,B, and \object{HD~95735}/GJ~411) at 7.5--7.7\,mag, far from the limit of the naked human eye. 

The specific advantage of CARMENES is the wide wavelength coverage and high spectral resolution in both visible and near-infrared channels.
Simultaneous observation from 0.5 to 1.7\,$\mu$m is a powerful tool for distinguishing between genuine planet detections and false positives caused by stellar activity, which have plagued planet searches employing spectrographs with a smaller wavelength coverage, especially in the M-type spectral domain (Reiners et~al. 2010; Barnes et~al. 2011). 
A substantial amount of guaranteed time for the completion of the key project is also an asset.

A precise knowledge of the targets is critical to ensure that most of the CARMENES guaranteed time is spent on the most promising targets.
This selection involves not only a comprehensive data compilation from the literature, but also summarises our observational effort to achieve new low- and high-resolution optical spectroscopy and high-resolution imaging.
The present publication on low-resolution spectroscopy is the first paper of a series aimed at describing the selection and characterisation of the CARMENES sample.
We have shown some preliminary results at conferences before
that described the input catalogue description and selection (Caballero et~al. 2013; Morales et~al. 2013), low-resolution spectroscopy (Klutsch et~al. 2012; Alonso-Floriano et~al. 2013a), high-resolution spectroscopy (Alonso-Floriano et~al. 2013b; Passegger et~al. 2014), resolved multiplicity (B\'ejar et~al. 2012; Cort\'es-Contreras et~al. 2013), X-rays (Lalitha et~al. 2012), exploitation of public databases (Montes et~al. 2015), or synergies with {\em Kepler} K2 (Rodr\'{\i}guez-L\'opez et~al. 2014). 
This first item of the CARMENES science-preparation series details the low-resolution optical spectroscopy of M dwarfs with the CAFOS spectrograph at the Zeiss 2.2\,m Calar Alto telescope.

\section{CARMENES sample}
\label{section.sample}

\begin{table}[]
\centering
\caption{Completeness and limiting $J$-band magnitudes per spectral type for the CARMENES input catalogue.}
\label{table.carmencitalimits}
 \begin{tabular}{l c c} 
   \hline
   \hline
   \noalign{\smallskip}
Spectral type   &  \multicolumn{2}{c}{$J$ [mag]}        \\
   \noalign{\smallskip}
    \cline{2-3}
    \noalign{\smallskip}
                        & Completeness  & Limiting         \\
   \noalign{\smallskip}
    \hline
    \noalign{\smallskip}
M0.0--0.5\,V    & 7.3                   & 8.5                   \\
M1.0--1.5\,V    & 7.8                   & 9.0                   \\
M2.0--2.5\,V    & 8.3                   & 9.5                   \\
M3.0--3.5\,V    & 8.8                   & 10.0          \\
M4.0--4.5\,V    & 9.3                   & 10.5          \\
M5.0--5.5\,V    & 9.8                   & 11.0          \\
M6.0--6.5\,V    & 10.3                  & 11.5          \\
M7.0--7.5\,V    & 10.8                  & 11.5          \\
M8.0--9.5\,V    & 11.3                  & 11.5          \\
   \noalign{\smallskip}
\hline
\end{tabular}   
\end{table}

\begin{table*}[]
\centering
\caption{Sources of the CARMENES input catalogue.}
\label{table.carmencitasources}
\begin{tabular}{l l c} 
   \hline
   \hline
   \noalign{\smallskip}
Source                                                                                                                  & Reference$^{a}$                 & Number                \\
                                                                                                                                &                                                 & of stars              \\
   \noalign{\smallskip}
    \hline
    \noalign{\smallskip}
The Palomar/MSU nearby star spectroscopic survey                                                & PMSU$^{b}$                      & 676   \\
A spectroscopic catalog of the brightest ($J <$ 9) M dwarfs in the northern sky                & L\'epine et~al. 2013          & 446     \\
G. P. Kuiper's spectral classifications of proper-motion stars                                  & Bidelman 1985                   & 285   \\
An all-sky catalog of bright M dwarfs$^{c}$                                                             & L\'epine \& Gaidos 2011 & 248   \\
Spectral types of M dwarf stars                                                                                 & Joy \& Abt 1974                 & 223   \\
Spectral classification of high-proper-motion stars                                                     & Lee 1984                                & 118   \\
Meeting the cool neighbors                                                                                      & RECONS$^{d}$                    & 22            \\
Search for nearby stars among proper... III. Spectroscopic distances of 322 NLTT stars & Scholz et~al. 2005          & 19            \\
New neighbors: parallaxes of 18 nearby stars selected from the LSPM-North catalog & L\'epine et~al. 2009               & 13            \\
Near-infrared metallicities, radial velocities and spectral types for 447 nearby M dwarfs & Newton et~al. 2014    & 10            \\
   \noalign{\smallskip}
\hline
\end{tabular}   
\begin{list}{}{}
\item[$^{a}$] Some other publications and meta-archives that we have searched for potential CARMENES targets are
Kirkpatrick et~al. (1991),
Gizis (1997),
Gizis \& Reid (1997),
Gizis et~al. (2000b),
Henry et~al. (2002, 2006),
Mochnaki et~al. (2002),
Gray et~al. (2003),
Bochanski et~al. (2005),
Crifo et~al. (2005),
Lodieu et~al. (2005),
Scholz et~al. (2005),
Phan-Bao \& Bessell (2006),
Reyl\'e et~al. (2006),
Riaz et~al. (2006),
Caballero (2007, 2009, 2012), 
Gatewood \& Coban (2009),
Shkolnik et~al. (2009, 2012),
Bergfors et~al. (2010),
Johnson et~al. (2010),
Boyd et~al. (2011),
Irwin et~al. (2011),
West et~al. (2011),
Avenhaus et~al. (2012),
Deacon et~al. (2012),
Janson et~al. (2012, 2014),
Frith et~al (2013),
J\'odar et~al. (2013),
Malo et~al. (2013),
Aberasturi et~al. (2014), 
Dieterich et~al. (2014),
Riedel et~al. (2014),
Yi et~al. (2014),
Gaidos et~al. (2014),
and the DwarfArchive at {\tt http://dwarfarchive.org}.
\item[$^{b}$] PMSU: 
Reid et~al. 1995, 2002; 
Hawley et~al. 1996; 
Gizis et~al. 2002.
\item[$^{c}$] With spectral types derived from spectroscopy in this work.
\item[$^{d}$] RECONS: 
Henry et~al. 1994; 
Kirkpatrick et~al. 1995; 
Henry et~al. 2006;
Jao et~al. 2011;
Riedel et~al. 2014; 
Winters et~al. 2015 and references therein.
\end{list}
\end{table*}

To prepare the CARMENES input catalogue with the best targets, we systematically collected all published M dwarfs in the literature that fulfilled two simple criteria:
\begin{itemize}
\item They had to be observable from Calar Alto with target declinations $\delta >$ --23\,deg (i.e., zenith distances $<$ 60\,deg, air masses at culmination $<$ 2.0).
\item They were selected according to late spectral type and brightness.
We only catalogued confirmed dwarf stars with an accurate spectral type determination from spectroscopic data (i.e., not from photometry) between M0.0\,V and M9.5\,V.
Additionally, we only compiled the brightest stars of each spectral type.
Our database contains virtually all known M dwarfs that are brighter than the completeness magnitudes shown in Table~\ref{table.carmencitalimits}, and most of them brighter than the limiting magnitudes. 
No target fainter than $J$ = 11.5\,mag is in our catalogue.
\end{itemize}

We started to fill the CARMENES database with the M dwarfs from the Research Consortium on Nearby Stars at {\tt http://www.recons.org}, which catalogues all known {stars with measured astrometric parallaxes that place them within 10\,pc} (e.g., 
Henry et~al. 1994; 
Kirkpatrick et~al. 1995; 
Riedel et~al. 2014; 
Winters et~al. 2015).
The RECONS {stellar compilation} was next completed with the Palomar/Michigan State University survey catalogue of nearby stars (PMSU -- Reid et~al. 1995, 2002; Hawley et~al. 1996; Gizis et~al. 2002).
Afterwards, we gave special attention to the comprehensive proper-motion catalogues of L\'epine et~al. (2003, 2009, 2013) and L\'epine \& Gaidos (2011), and the ``Meeting the Cool Neighbors'' series of papers 
(Cruz \& Reid 2002; 
Cruz et~al. 2003,       
2007;                   
Reid et~al. 2003,       
2004,                   
2008).                  
Table~\ref{table.carmencitasources} provides the sources of our
information on M dwarfs.
Until we start our survey at the end of 2015, we will still include some new, particularly bright, late, single, M dwarfs\footnote{Please contribute to the comprehensiveness of our input catalogue by sending an e-mail with suggestions to Jos\'e A. Caballero, \email{caballero@cab.inta-csic.es}.}.

As of February 2015, our input catalogue, dubbed CARMENCITA (CARMENES Cool dwarf Information and daTa Archive), contains approximately 2200 M dwarfs.
For each target star, we tabulate a number of parameters  compiled from the literature or measured by us with new data: 
accurate astrometry and distance, spectral type, photometry in 20 bands from the ultraviolet to the mid-infrared, rotational, radial, and Galactocentric velocities, H$\alpha$ emission, X-ray count rates and hardness ratios, close and wide multiplicity data, membership in open clusters and young moving groups, target in other radial-velocity surveys, and exoplanet candidacy (Caballero et~al. 2013). 
The private on-line catalogue, including preparatory science observations (i.e., low- and high-resolution spectroscopy, high-resolution imaging), will become public as a CARMENES legacy. 

Of the 2200 stars, we discard all spectroscopic binaries and multiples, and resolved systems with physical or visual companions at less than 5\,arcsec to our targets.
The size of the CARMENES optical fibres projected on the sky is 1.5\,arcsec (Seifert et~al. 2012; Quirrenbach et~al. 2014), and consequently any companion at less than 5\,arcsec may induce real or artificial radial-velocity variations that would contaminate our measurements (Guenther \& Wuchterl 2003; Ehrenreich et~al. 2010; Guenther \& Tal-Or 2010).
About 1900 single stars currently remain after discarding all multiple systems.

\begin{figure*}
\centering
\includegraphics[trim= 40 0 0 0,width=1.00\textwidth]{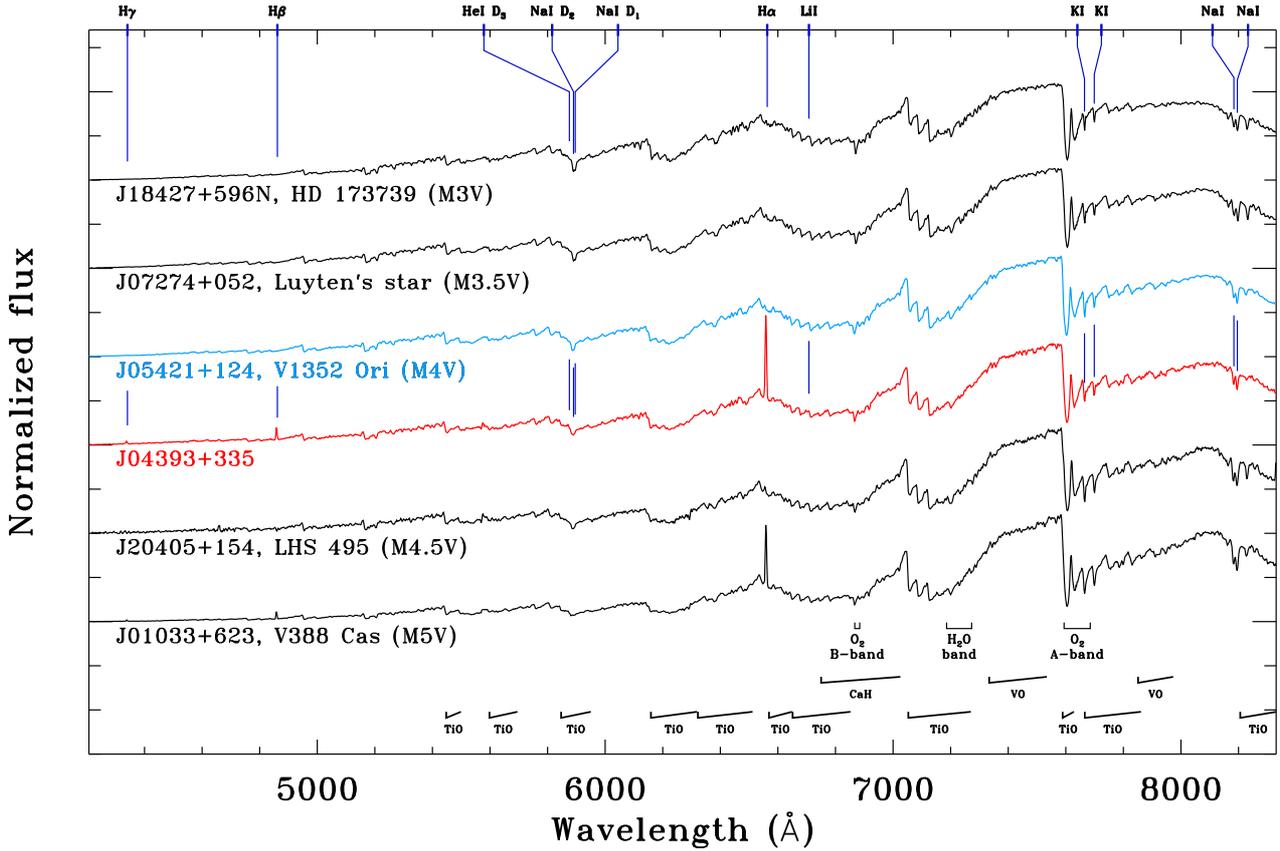}
\caption{Six representative CAFOS spectra.
From top to bottom, spectra of
standard stars with spectral type 1.0 and 0.5 subtypes earlier than the target (black),
standard star with the same spectral type as the target (cyan),
the target star (red; in this case, J04393+335 -- M4.0\,V, Simbad name: \object{V583~Aur\,B}),
and standard stars with spectral type 0.5 and 1.0 subtypes later than the target (black).
We mark activity-, gravity-, and youth-sensitive lines and doublets at the top of the figure (H$\gamma$, H$\beta$, He~{\sc i}~D$_3$, Na~{\sc i}~D$_2$ and D$_1$, H$\alpha$, Li~{\sc i}, K~{\sc i}, and Na~{\sc i}, from left to right) and molecular absorption bands at the bottom.
Note the three first lines of the Balmer series in emission in the spectrum of the target star.} 
\label{fig.classification}
\end{figure*}

\subsection{CAFOS sample}

The aim of our low-resolution spectroscopic observations is twofold:
$(i)$ to increase the number of bright, late-M dwarfs in CARMENCITA and $(ii)$ to ensure that the compiled spectral types used for the selection and pre-cleaning are correct.
With this double objective in mind, we observed the following:

\begin{itemize}

\item High proper-motion M-dwarf candidates from L\'epine \& Shara (2005) and L\'epine \& Gaidos (2011) with spectral types with large uncertainties or derived only from photometric colours.
Spectral types from $V^*-J$ colours are not suitable for our purposes (Alonso-Floriano et~al. 2013a; Mundt et~al. 2013; L\'epine et~al. 2013 -- $V^*$ is an average of photographic magnitudes $B_J$ and $R_F$ from the Digital Sky Survey; cf. L\'epine \& Gaidos 2011).
In collaboration with  Sebastien~L\'epine, we observed and analysed an extension of the L\'epine \& Gaidos (2011) catalogue of high proper-motion candidates brighter than $J$ = 10.5\,mag.
The spectra of stars brighter than $J$ = 9.0\,mag were published by L\'epine et~al. (2013), while most of the remaining fainter ones are published here.

\item M dwarf candidates in nearby young moving groups (e.g., Montes et~al. 2001; Zuckermann \& Song 2004; da~Silva et~al. 2009; Shkolnik et~al. 2012; Gagn\'e et~al. 2014; Klutsch et~al 2014), in multiple systems containing FGK-type primaries that
are subjects of metallicity studies (Gliese \& Jahreiss 1991; Poveda et~al. 1994; Gould \& Chanam\'e 2004; Rojas-Ayala et~al. 2012; Terrien et~al. 2012; Mann et~al. 2013; Montes et~al. 2013), in fragile binary systems at the point of disruption by the Galactic gravitational field (Caballero 2012 and references therein), and resulting from new massive virtual-observatory searches (Jim\'enez-Esteban et~al. 2012; Aberasturi et~al. 2014).
Such a broad diversity of sources allowed us to widen the investigated intervals of age, activity, multiplicity, metallicity, and dynamical evolution.

\item Known M dwarfs with well-determined spectral types from PMSU (see above) and L\'epine et~al. (2013).
The comparison of these two samples with ours was a sanity check for determining the spectral types (see Sect.~\ref{subsection:SpT}).

\item M dwarfs in our input catalogue with uncertain or probably incorrect spectral types based on apparent magnitudes, $r'-J$ colours, and heliocentric distances, including resolved physical binaries.
See some examples in Cort\'es-Contreras et~al. (2014).

\item Numerous standard stars.
For an accurate determination of spectral type and class, we also included approximately 50 stars with well-determined spectral types from K3 to M8 for both dwarf (Johnson \& Morgan 1953; Kirkpatrick et~al. 1991; PMSU) and giant classes (e.g., Moore \& Paddock 1950; Ridgway et~al. 1980; Jacoby et~al. 1984; Garc\'{\i}a 1989; Keenan \& McNeil 1989; Kirkpatrick et~al. 1991; S\'anchez-Bl\'azquez et~al. 2006; Jim\'enez-Esteban et~al. 2012).

\end{itemize}

\section{Observations and analysis}

\subsection{Low-resolution spectroscopic data}

Observations were secured with the Calar Alto Focal reductor and Spectrograph (CAFOS) mounted on the Ritchey-Chr\'etien focus of the Zeiss 2.2\,m Calar Alto telescope (Meisenheimer 1994).
We obtained more than 900 spectra of {745} targets during 38 nights over four semesters from 2011 November to 2013 April. 
All observations were carried out in service mode with the G-100 grism, which resulted in a useful wavelength coverage of 4200--8300\,{\AA} at a resolution $\mathcal{R} \sim$ 1500.
Exposure times ranged from shorter than 1\,s to 1\,h.
The longest exposures were split into up to four sub-exposures.
On some occasions, another star fell in the slit aperture (usually the primary of a close multiple system containing our M-dwarf candidate main target).
We also added the {13} red dwarfs and giants observed in 2011 March by Jim\'enez-Esteban et~al. (2012), which made a total of {758} targets.
 
\begin{table}[]
\centering
\caption{Standard and prototype stars$^{a}$.}
\label{table.standards}
\begin{tabular}{l c l l} 
   \hline
   \hline
   \noalign{\smallskip}
SpT             &       & Karmn         &       Name    \\
   \noalign{\smallskip}
    \hline
    \noalign{\smallskip}
M0.0\,V         & *     &        J09143+526      &   HD 79210                   \\
                &       &        J07195+328      &   BD+33~1505         \\
                &       &        J09144+526      &   HD 79211                   \\
M0.5\,V         & *     &   J18353+457   &   BD+45~2743         \\ 
                &       &    J04329+001S         &   LP 595--023                \\
                &       &    J22021+014          &   HD 209290                  \\ 
M1.0\,V         & *     &        J00183+440      &   GX And                     \\ 
                &       &        J05151--073     &   LHS 1747                   \\
                &       &        J11054+435      &   BD+44~2051A                \\ 
M1.5\,V         & *     &    J05314--036         &   HD 36395                   \\
                &       &    J00136+806          &   G~242--048         \\
                &       &    J01026+623          &   BD+61~195          \\
                &       &    J11511+352          &   BD+36~2219         \\
M2.0\,V         & *     &        J08161+013      &   GJ 2066                    \\
                &       &        J03162+581N     &   Ross 370 B                       \\
                &       &        J03162+581S     &   Ross 370 A                       \\
M2.5\,V         & *     &    J11421+267          &   Ross 905                   \\ 
                &       &    J10120--026 AB      &   LP 609--071                \\
                &       &    J19169+051N         &   V1428 Aql                  \\
                &       &    J21019--063         &   Wolf 906                   \\
M3.0\,V         & *     &        J18427+596N     &   HD 173739                  \\
                &       &        J17364+683      &   BD+68~946          \\  
                &       &        J22524+099      &   $\sigma$~Peg B               \\
M3.5\,V         & *     &    J07274+052          &    \object{Luyten's star}           \\
                &       &    J17199+265          &    \object{V647 Her}            \\
                &       &    J17578+046          &    \object{Barnard's star}   \\
                &       &    J18427+596S         &    \object{HD 173740}         \\
M4.0\,V         & *     &        J05421+124      &   \object{V1352 Ori}            \\
                &       &        J04308--088     &   \object{Koenigstuhl 2 A}    \\
                &       &        J06246+234      &   \object{Ross 64}                     \\
                &       &        J10508+068      &   \object{EE Leo}                    \\ 
M4.5\,V         & *     &    J20405+154          &   \object{G 144--025}               \\
                &       &        J04153--076     &   \object{$o^{02}$ Eri C}  \\ 
                &       &    J16528+610          &   \object{GJ 625}                    \\
                &       &    J17198+265          &   \object{V639 Her     }               \\ 
M5.0\,V         & *     &        J01033+623      &   \object{V388 Cas}            \\
                &       &        J16042+235      &   \object{LSPM J1604+2331} \\
                &       &        J23419+441      &   \object{HH And}                    \\
M5.5\,V         & *     &  J02022+103    &   \object{LP 469--067}               \\
                &       &    J21245+400          &   \object{LSR J2124+4003} \\
M6.0\,V         & *     &        J10564+070      &   \object{CN Leo}                    \\
                &       &        J07523+162      &   \object{LP 423--031}               \\ 
                &       &        J16465+345      &   \object{LP 276--022}               \\ 
M6.5\,V         & *     &    J08298+267          &   \object{DX Cnc}                    \\
                &       &    J09003+218          &   \object{LP 368--128}               \\
                &       &    J10482--113         &   \object{LP 731--058}               \\
M7.0\,V         & *     &        J16555-083      &   V1054 Oph D (\object{vB 8}) \\
                &       &    J02530+168          &   \object{Teegarden's star}   \\
M8.0\,V         & *     &        J19169+051S     &   \object{V1298 Aql} (vB\,10)   \\
\noalign{\smallskip}
\hline
\end{tabular}   
\begin{list}{}{}
\item[$^{a}$] Prototype stars are marked with an asterisk.
\end{list}
\end{table}

\begin{figure}[]
\includegraphics[trim=30 0 0 0, width=0.49\textwidth]{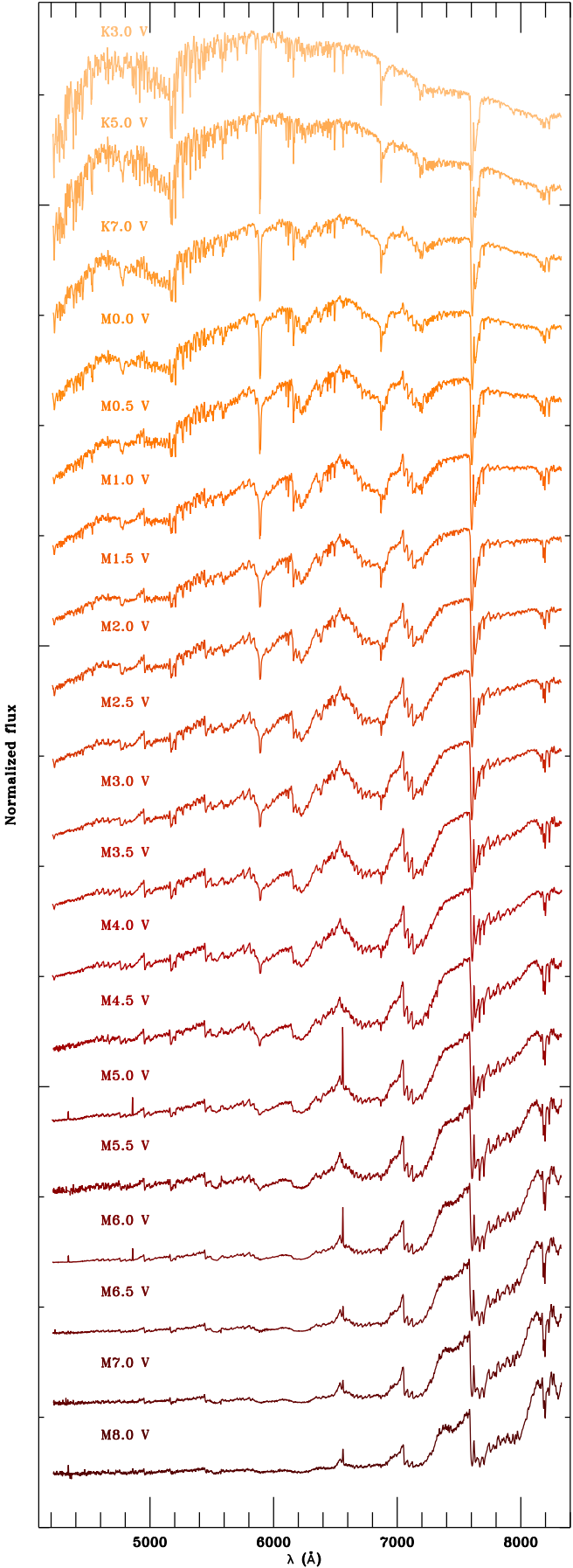}
 \caption{\label{figure.prototypes} CAFOS spectra of our prototype stars.
 From top to bottom, K3\,V, K5\,V, K7\,V, M0.0--7.0\,V in steps of 0.5 subtypes, and M8.0\,V.} 
\end{figure}

We reduced the spectra using typical tasks within the IRAF environment.
The reduction included bias subtraction, flat fielding, removal of sky background, optimal aperture extraction, wavelength calibration (with Hg-Cd-Ar, He, and Rb lamps), and instrumental response correction. 
For the latter, we repeatedly observed the spectrophotometric standards G\,191--B2B (DA0.8), HD~84937 (sdF5), Feige~34 (sdO), BD+25~3941 (B1.5\,V), and  BD+28~421 (sdO) at different air masses. 
In the end, we only used the spectra with the highest signal-to-noise ratio of the hot subdwarf \object{Feige~34}, which gave the best-behaved instrumental response correction. 
We extracted all traces in the spectra, including those of other stars in the slit aperture.
We did not remove telluric absorption lines from the spectra that were due to the variable meteorological conditions during two years of observation (see Sect.~\ref{subsec:least_square_minimisation}). 
All our spectra had a signal-to-noise ratio higher than 50 near the H$\alpha$~$\lambda$6562.8\,{\AA} line, which together with the wide wavelength coverage allowed us to make a comprehensive analysis and to measure numerous spectral indices and activity indicators.

We list in Table~\ref{table.observed}  the {753} observed K and M dwarf and giant candidates according to identification number, our CARMENCITA identifier, discovery name, {Gliese or Gliese \& Jahrei{\ss} number}, J2000.0 coordinates and $J$-band magnitude from the Two-Micron All-Sky Survey (Skrutskie et~al. 2006), observation date, and exposure time. 
The {five} stars not tabulated are the spectrophotometric standards.

In Table~\ref{table.observed}, our CARMENCITA identifier follows the nomenclature format `Karmn~JHHMMm$\pm$DDd(X)', where `Karmn' is the acronym, `m' and `d' in the sequence are the truncated decimal parts of a minute or degree of the corresponding equatorial coordinates for the standard equinox of J2000.0 ({\em IRAS} style for right ascension, PKS quasar style for declination), and X is an optional letter (N, S, E, W) to distinguish between physical or visual pairs with the same HHMMm$\pm$DDd sequence within CARMENCITA.
We use the discovery name for every target, except for M dwarfs with variable names (e.g., EZ~PSc, GX~And, V428~And) or those that are physical companions to bright stars (e.g., BD--00~109\,B, $\eta$~Cas\,B, HD~6440\,B).
We associate for the first time many X-ray events with active M dwarfs (e.g., Lalitha et~al. 2012; Montes et~al. 2015). 
In these cases, we use the precovery\footnote{``Pre-discovery recovery''.} {\em Einstein}~2E or {\em ROSAT}~RX/1RXS event identifications instead of the names given by the proper-motion survey that recovered the stars.

Additionally, we always indicate whether the star is a known close binary or triple unresolved in our spectroscopic data (with `AB', 'BC', 'ABC'). 
There are {75} such close multiple systems in our sample ({70} double and {5} triple), of which the widest unresolved pair is J09045+164\,AB (\object{BD+16 1895}), with $\rho \approx$ 4.1\,arcsec.
A comprehensive study on the multiplicity of M dwarfs in CARMENCITA, including spectral-type estimate of companions from magnitude differences in high-resolution imaging data, will appear in a forthcoming paper of this series (preliminary data were presented in  Cort\'es-Contreras et~al. 2015).

\subsection{Spectral typing}

We followed two widely used spectral classification schemes that require an accurate, wide grid of reference stars.
The first strategy relies on least-squares minimisation and best-fitting to spectra of the standard stars, while the second scheme uses spectral indices that quantify the strength of the main spectral features in M dwarfs, notably molecular absorption bands (again, preliminary data were presented in Klutsch et~al. 2012 --least-squares minimisation-- and Alonso-Floriano et~al. 2013a --spectral indices).

Before applying the two spectral-typing strategies, we normalised our spectra by dividing by the observed flux at 7400\,{\AA}. 
We also corrected for spatial distortions at the reddest wavelengths (with R $\sim$ 1500, there is no need for a stellar radial-velocity correction).
For that, we shifted our spectra until the K~{\sc i} $\lambda\lambda$7664.9,7699.0\,{\AA} doublet was placed at the laboratory wavelengths. 
This correction, often of 1--2\,{\AA}, was critical for the definition of spectral indices, some of which are very narrow.

\subsubsection{Spectral standard and prototype stars}

We list the used standard stars in Table~\ref{table.standards} (see also Sect.~\ref{subsec:spt}), most of which were taken from Kirkpatrick et~al. (1991) and PMSU (Table~\ref{table.carmencitasources}).
Our intention was to provide one prototype star and up to four reference stars per half subtype, but this was not always possible, especially at the latest spectral types.
The prototype stars, shown in Fig.~\ref{figure.prototypes}, are the brightest, least active reference stars that have spectra with the highest signal-to-noise ratio, and that do not deviate significantly from the general trend during fitting.
In Table~\ref{table.standards}, the first star of each subtype is the prototype for that subtype. 
In the case of standard stars with different reported spectral types in the bibliography (with maximum differences of 0.5 subtypes), we chose the value that gave us less scatter in our fits.

We also used three K dwarfs from Kirkpatrick et~al. (1991), not listed in Table~\ref{table.standards}, to extend our grid of standard stars towards warmer effective temperatures.
The three K-dwarf standard stars are \object{HD~50281} (K3\,V), \object{61~Cyg~A} (K5\,V), and \object{$\eta$~Cas~B} (K7\,V).
The standard star LP~609--071\,AB is a close binary composed of an M2.5\,V star and a faint companion separated by $\rho$ 0.18$\pm$0.02\,arcsec (Delfosse et~al. 1999).
With $\Delta K$ = 0.95$\pm$0.05\,mag, the faint companion flux barely affects the primary spectrum in the optical.

\begin{figure}[]
\includegraphics[width=0.49\textwidth]{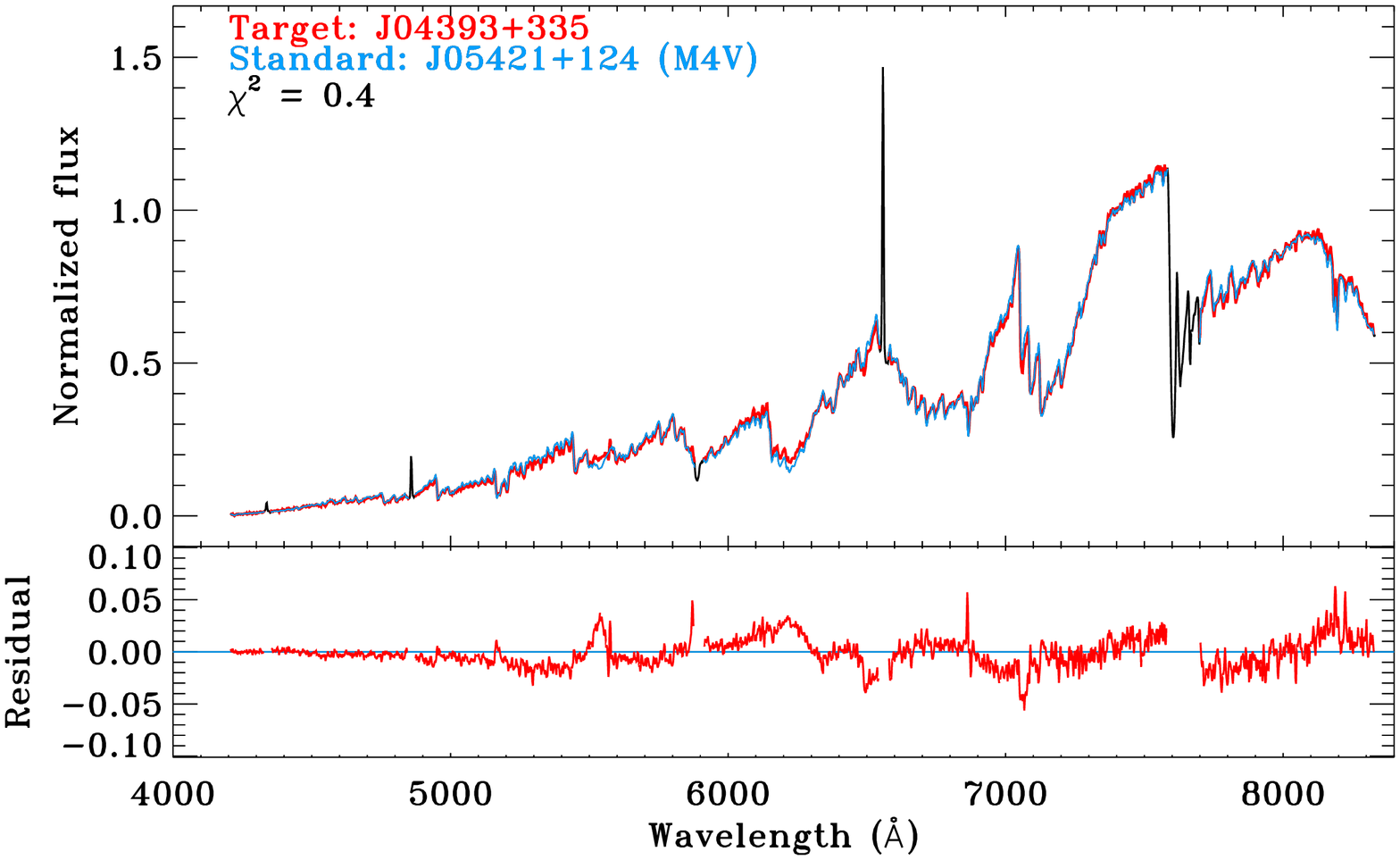}
\includegraphics[width=0.49\textwidth]{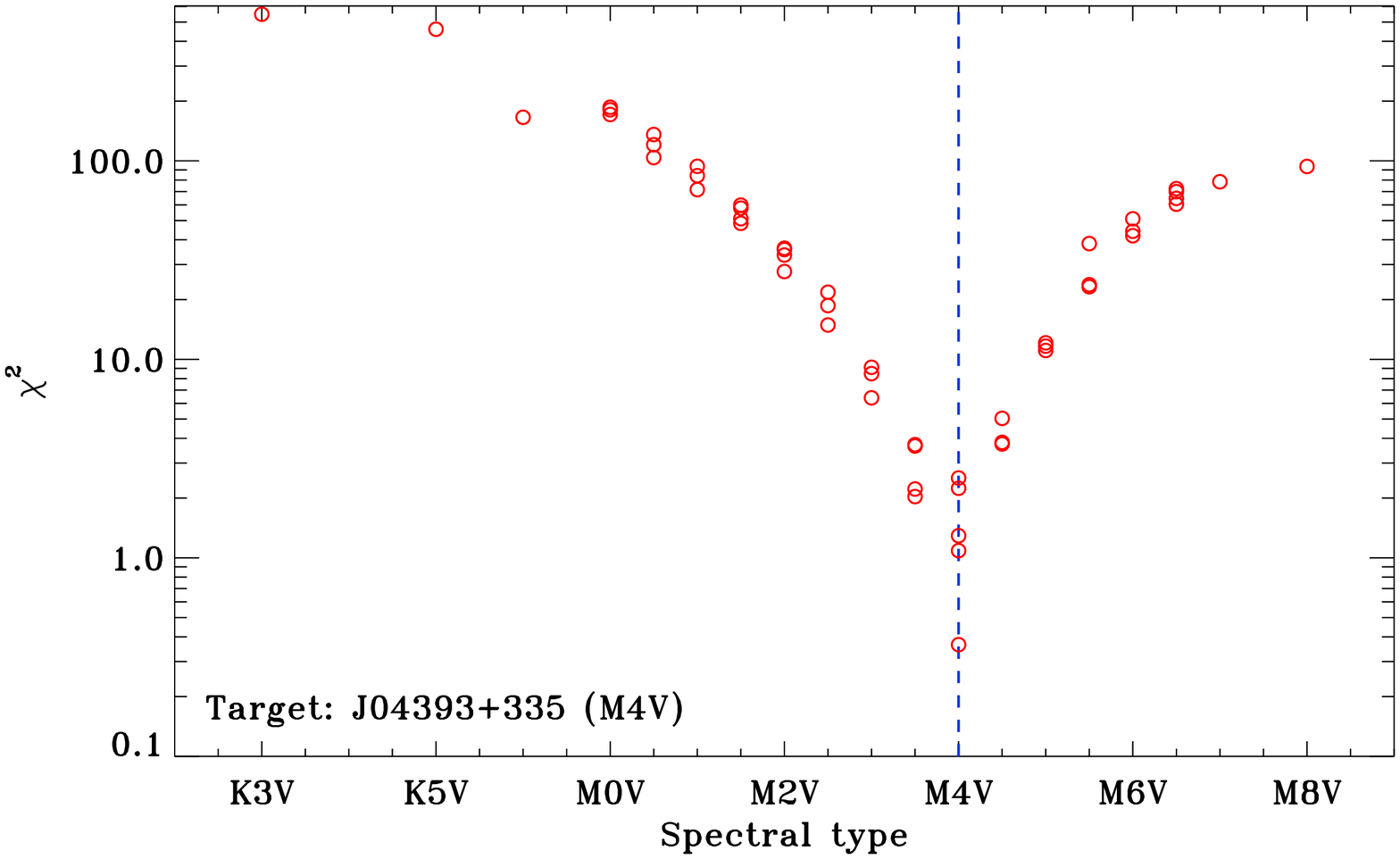}
\caption{\label{fig.bestmatch} Spectral typing of J04393+335 with best-match and $\chi^{2}_{\rm min}$ methods. 
\textit{Top panel:} best-match. 
CAFOS normalised spectra of the target star and of the standard star that fits best (top) and the difference (bottom).
\textit{Bottom panel:} $\chi^{2}_{\rm min}$.
Values of $\chi^{2}$ as a function of the spectral type (open red circles). 
The vertical dashed line marks the spectral type at the lowest $\chi^{2}$ value.
Note the logarithmic scale in the Y axis.}
\end{figure}

\subsubsection{Best-match and $\chi^{2}_{\rm min}$ methods}
\label{subsec:least_square_minimisation}

Before any least-squares minimisation, we discarded five narrow (20--30\,\AA) wavelength ranges influenced by activity indicators (H$\alpha$, H$\beta$, H$\gamma$, and the Na~{\sc i} doublet -- Fraunhofer C, F, and G' lines) and by strong telluric lines (O$_2$ band around 7594\,{\AA} -- Fraunhofer A line). 
Next, in the full remaining spectral range, we compared the normalised spectrum of every target star with those of all our standard stars in Table~\ref{table.standards} and computed a $\chi^2$ value for each fit.
In the best-match method, we assigned the spectral type of the standard star that best fitted our target spectrum (i.e., with the lowest $\chi^2$ value);
in the $\chi^{2}_{\rm min}$ method, we assigned the spectral type that corresponded to the minimum of the curve resulting from the (sixth-order) polynomial fit of all the $\chi^2$--spectral type pairs.
As expected, the best-match and $\chi^{2}_{\rm min}$ methods give the same result in most cases.
In the representative case shown in Figs.~\ref{fig.classification} and~\ref{fig.bestmatch}, the target dwarf has an M4.0\,V spectral type using both the best match and $\chi^{2}_{\rm min}$ methods.

\subsubsection{Spectral indices}
\label{subsec:spectral_indices}

\begin{table*}
        \centering
        \caption{Spectral indices used in this work.}
         \label{table.index}
        \begin{tabular}{ l c c l}
         \hline
         \hline
         \noalign{\smallskip}
Index           & $\Delta \lambda_{\rm num}$            &       $\Delta \lambda_{\rm dem}$      & Reference                                     \\
                        & [\AA]                                         &       [\AA]                                            &                                                       \\
 \noalign{\smallskip}           
          \hline
         \noalign{\smallskip}
CaOH            & 6230:6240                                     &       6345:6354                               &       {Reid et~al. 1995}                    \\
CaH\,1          & 6380:6390                                     &       $\sum$ 6345:6355, 6410:6420    &       {Reid et~al. 1995}                      \\
I2 (CaH)                & 6510:6540                                     &       6370:6400                               &       {Mart\'{\i}n \& Kun 1996}            \\ 
I3 (TiO)                & 6510:6540                                     &       6660:6690                               &       {Mart\'{\i}n \& Kun 1996}            \\ 
H$\alpha$       & 6560:6566                                     &       6545:6555                               &       {Reid et~al. 1995}                    \\ 
TiO\,1          & 6718:6723                                     &       6703:6708                               &       {Reid et~al. 1995}                    \\
CaH\,2          & 6814:6846                                     &       7042:7046                               &       {Reid et~al. 1995}                    \\
CaH\,3          & 6960:6990                                     &       7042:7046                               &       {Reid et~al. 1995}                    \\
TiO-7053                & 7000:7040                                     &       7060:7100                               &       {Mart\'{\i}n et~al. 1999}            \\ 
Ratio A (CaH)   & 7020:7050                                     &       6960:6990                               &       {Kirkpatrick et~al. 1991}            \\ 
TiO-7140                & 7015:7045                                     &       7125:7155                               &       {Wilking et~al. 2005}                    \\ 
PC1                     & 7030:7050                                     &       6525:6550                               &       {Mart\'{\i}n et~al. 1996}            \\
CaH\,Narr               & 7044:7049                                     &       6972.5:6977.5                           &       {Shkolnik et~al. 2009}            \\ 
TiO\,2          & 7058:7061                                     &       7043:7046                               &       {Reid et~al. 1995}                    \\ 
TiO\,3          & 7092:7097                                     &       7079:7084                               &       {Reid et~al. 1995}                    \\
TiO\,5          & 7126:7135                                     &       7042:7046                               &       {Reid et~al. 1995}                    \\
TiO\,4          & 7130:7135                                     &       7115:7120                               &       {Reid et~al. 1995}                    \\
VO-a                    & $\sum$ 7350:7370, 7550:7570   &       7430:7470                                 &       {Kirkpatrick et~al. 1999}               \\ 
VO                      & $\sum$ $\alpha$7350:7400, $\beta$7510:7560$^{a}$      & 7420:7470       &       {Kirkpatrick et~al. 1995}               \\
Ratio B (Ti~{\sc i}) & 7375:7385                                &       7353:7363                               &       {Kirkpatrick et~al. 1991}            \\ 
VO-7434         & 7430:7470                                     &       7550:7570                               &       {Hawley et~al. 2002}                    \\ 
PC2                     & 7540:7580                                     &       7030:7050                               &       {Mart\'{\i}n et~al. 1996}            \\
VO\,1           & 7540:7580                                     &       7420:7460                               &       {Mart\'{\i}n et~al. 1999}            \\ 
TiO\,6          & 7550:7570                                     &       7745:7765                               &       {L\'epine et~al. 2003}                    \\
VO-b                    & $\sum$ 7860:7880, 8080:8100   &       7960:8000                               &       {Kirkpatrick et~al. 1999}            \\ 
VO~2            & 7920:7960                                     &       8130:8150                               &       {L\'epine et~al. 2003}                    \\
VO-7912         & 7990:8030                                     &       7900:7940                               &       {Mart\'{\i}n et~al. 1999}            \\ 
Ratio C (Na~{\sc i}) & 8100:8130                                &       8174:8204                               &       {Kirkpatrick et~al. 1991}            \\ 
Color-M         & 8105:8155                                     &       6510:6560                               &       {L\'epine et~al. 2003}                    \\
Na-8190         & 8140:8165                                     &       8173:8210                               &       {Hawley et~al. 2002}                    \\ 
PC3                     & 8235:8265                                     &         7540:7580                               &       {Mart\'{\i}n et~al. 1996}            \\ 
\noalign{\smallskip}
\hline
        \end{tabular}
\begin{list}{}{}
\item[$^{a}$] $\alpha$ = 0.5625, $\beta$ = 0.4375.
\end{list}
\end{table*}

The spectral indices methodology for spectral typing is based on computing flux ratios at certain wavelength intervals in low-resolution spectra (e.g., Kirkpatrick et~al. 1991; Reid et~al. 1994; Mart\'{\i}n et~al. 1996, 1999).
In the present analysis, we compiled 31 spectral indices defined in the literature to determine spectral types of late-K dwarfs and M dwarfs that occur in the useable wavelength interval of our CAFOS spectra.
In general, a spectral index $I_i$ is defined by the ratio of numerator and denominator fluxes (i.e, $I_i = F_{i,\rm num} / F_{i,\rm den}$).
Table~\ref{table.index} lists the 31 wavelength intervals of fluxes in the numerator and denominator, $\Delta \lambda_{\rm num}$ and $\Delta \lambda_{\rm den}$, and corresponding reference for each index.  
Some flux wavelength intervals are the linear combination of two subintervals (CaH\,1, VO-a, VO-b) or, in the case of the VO index, a nonlinear combination.
Additionally, there are wavelength intervals of fluxes in the numerator that are either redder and bluer than the one in the denominator, which translates into different slopes in the index--spectral type relations.
Of the 31 tabulated indices, nine are related to TiO features, seven to VO, six to CaH, three to the ``pseudo-continuum'' (i.e., relative absence of features), and the rest to H and neutral metallic lines (Ti, Na).

\begin{figure*}[]
\center
\includegraphics[width=0.32\textwidth]{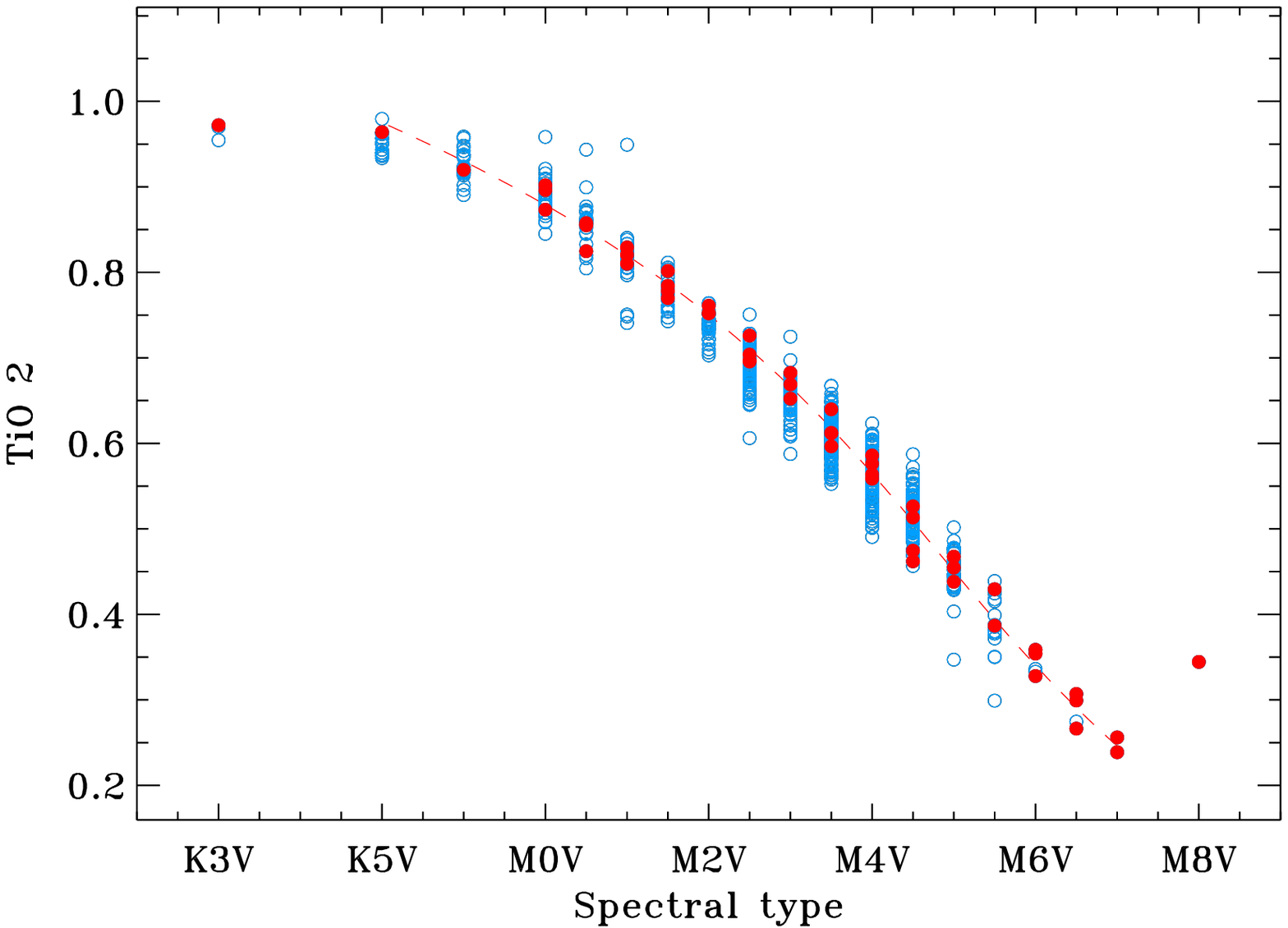}
\includegraphics[width=0.32\textwidth]{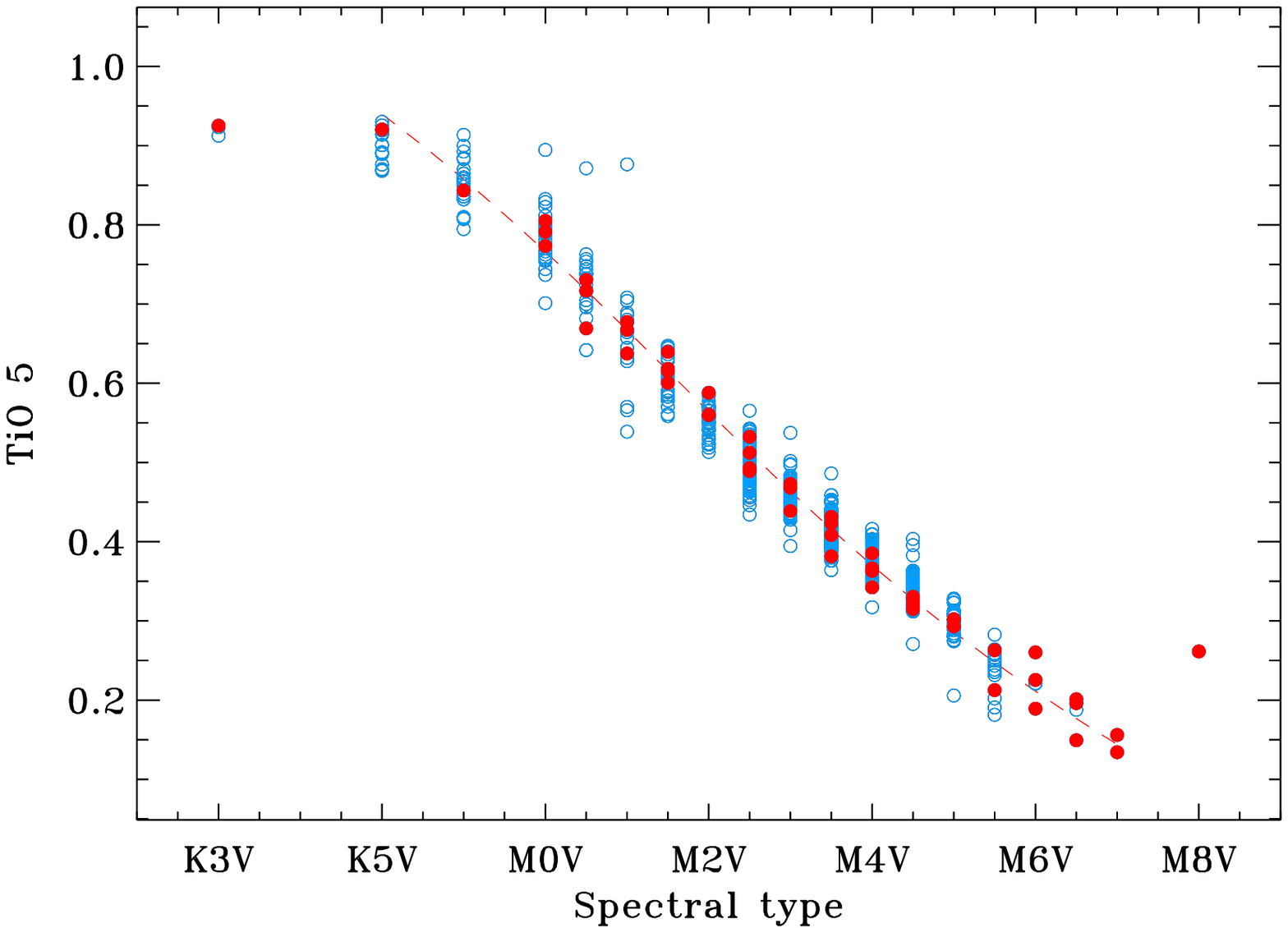}
\includegraphics[width=0.32\textwidth]{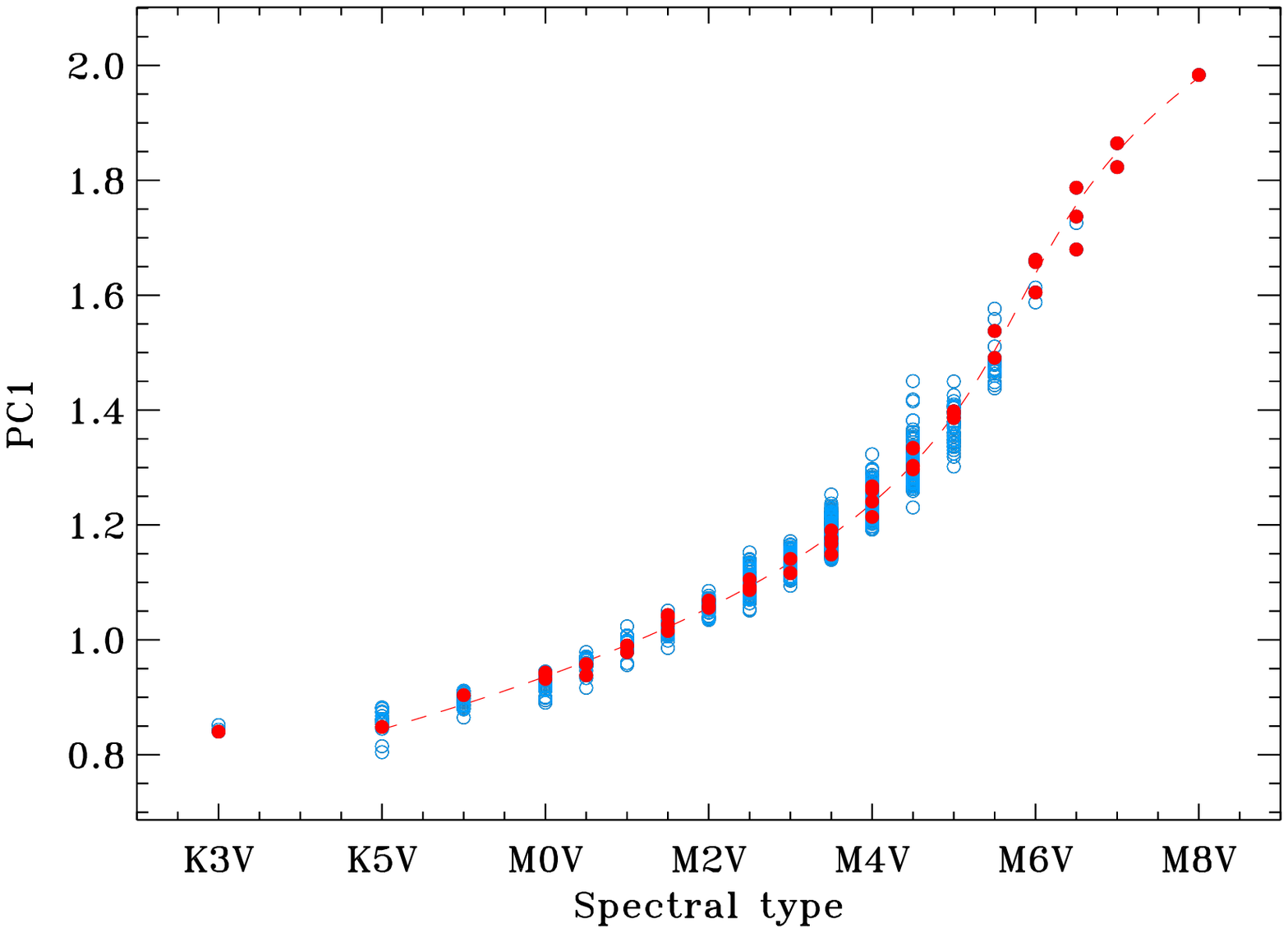}
\includegraphics[width=0.32\textwidth]{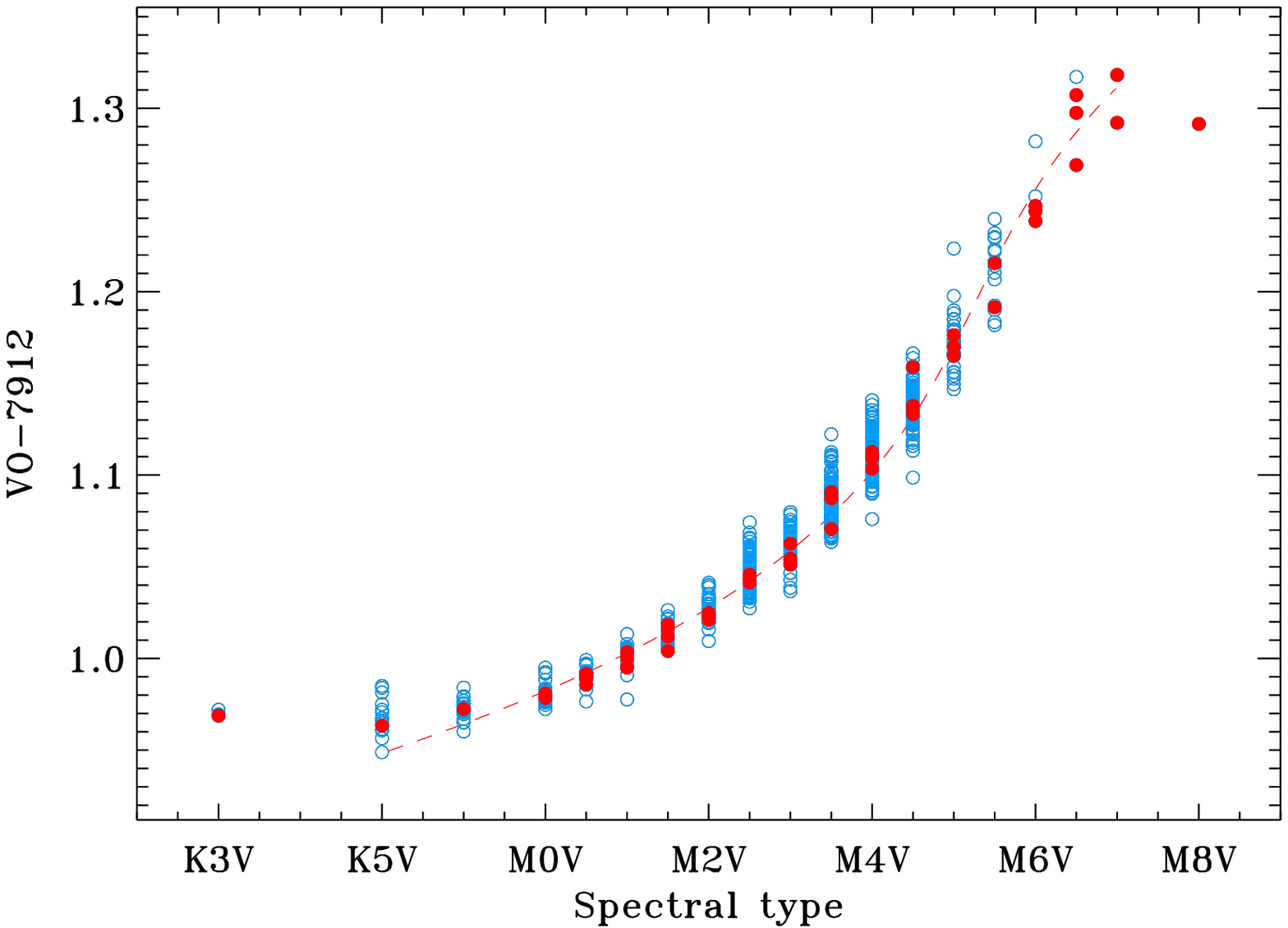}
\includegraphics[width=0.32\textwidth]{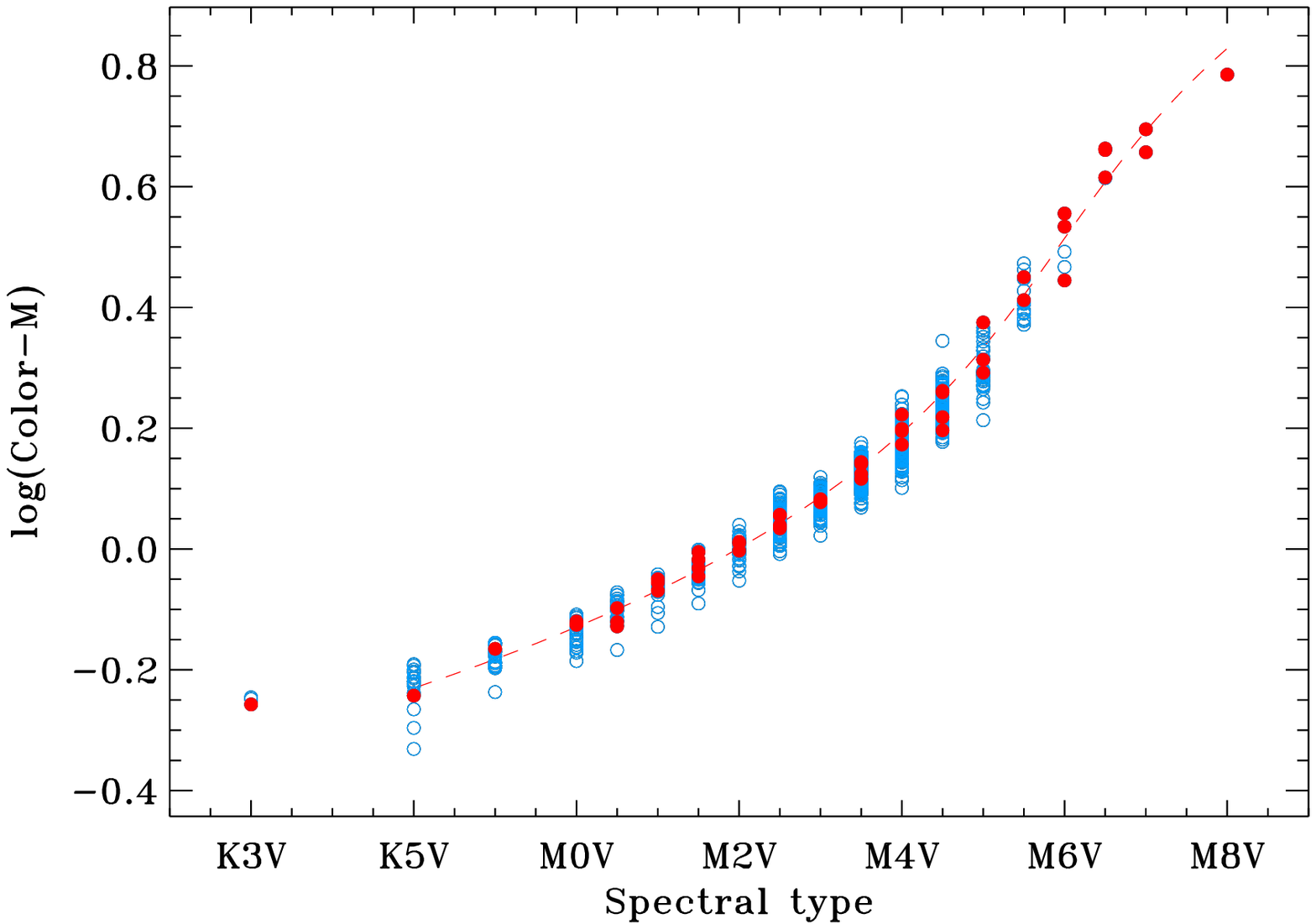}
\includegraphics[width=0.32\textwidth]{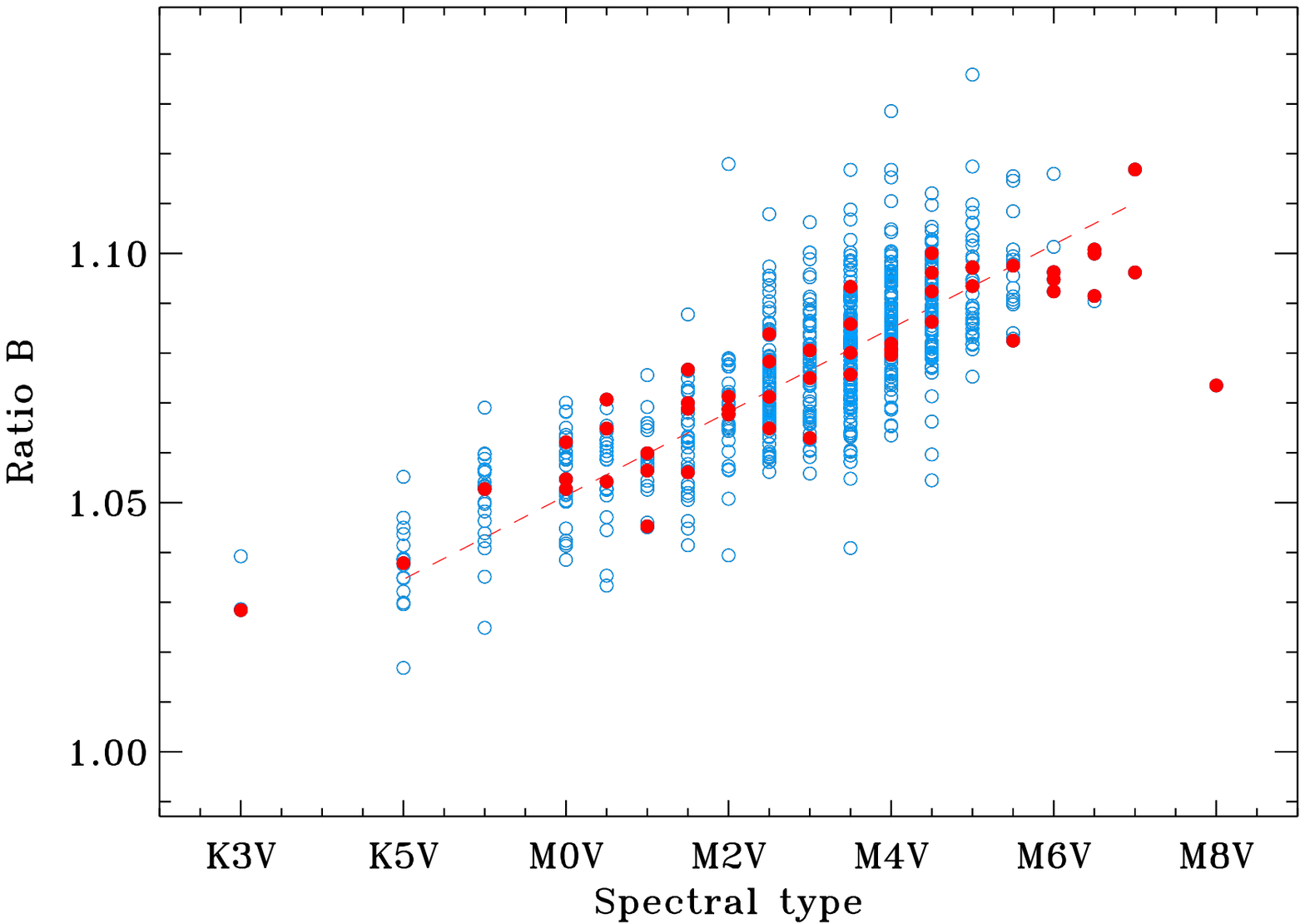}
\includegraphics[width=0.32\textwidth]{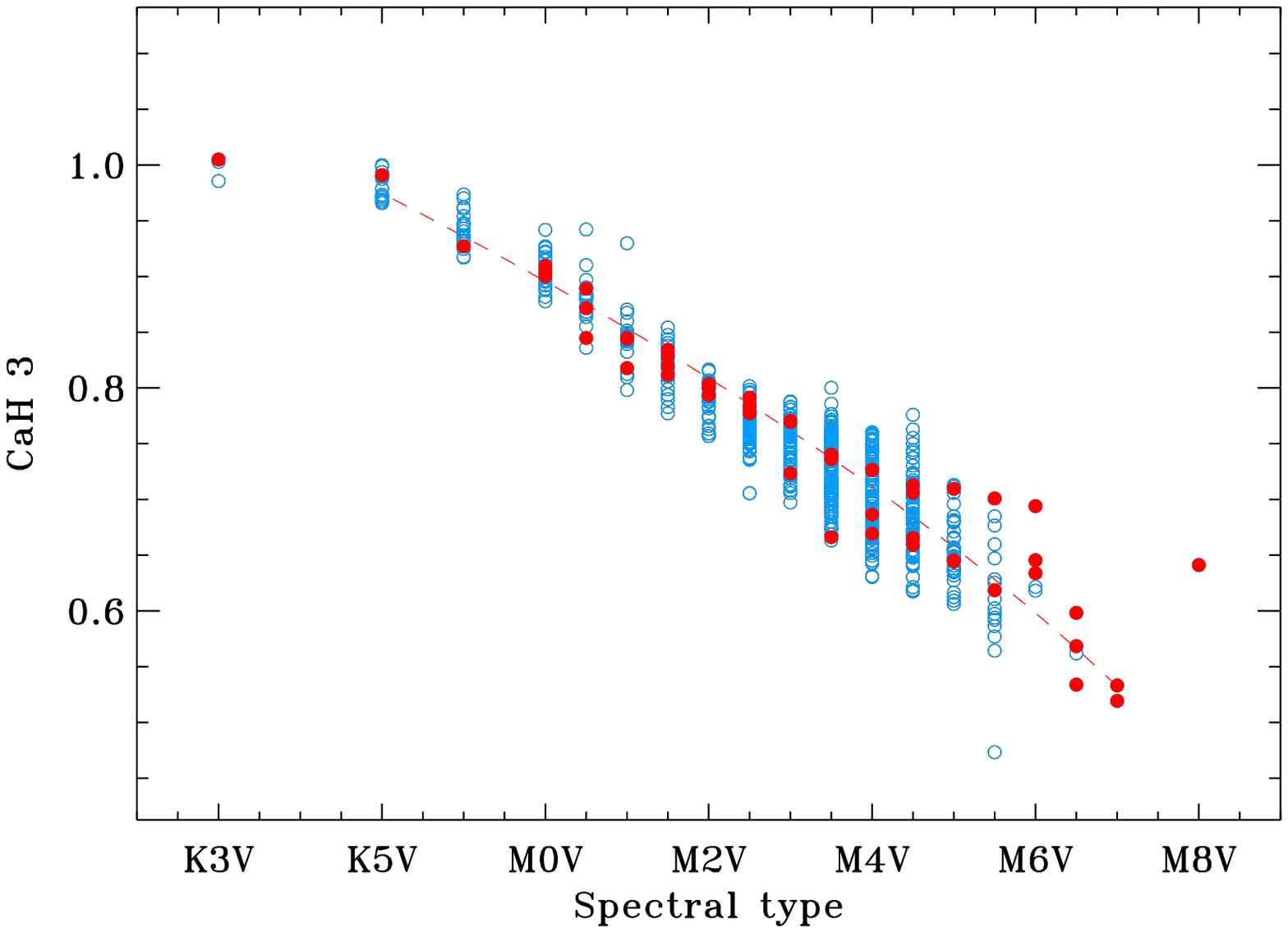}
\includegraphics[width=0.32\textwidth]{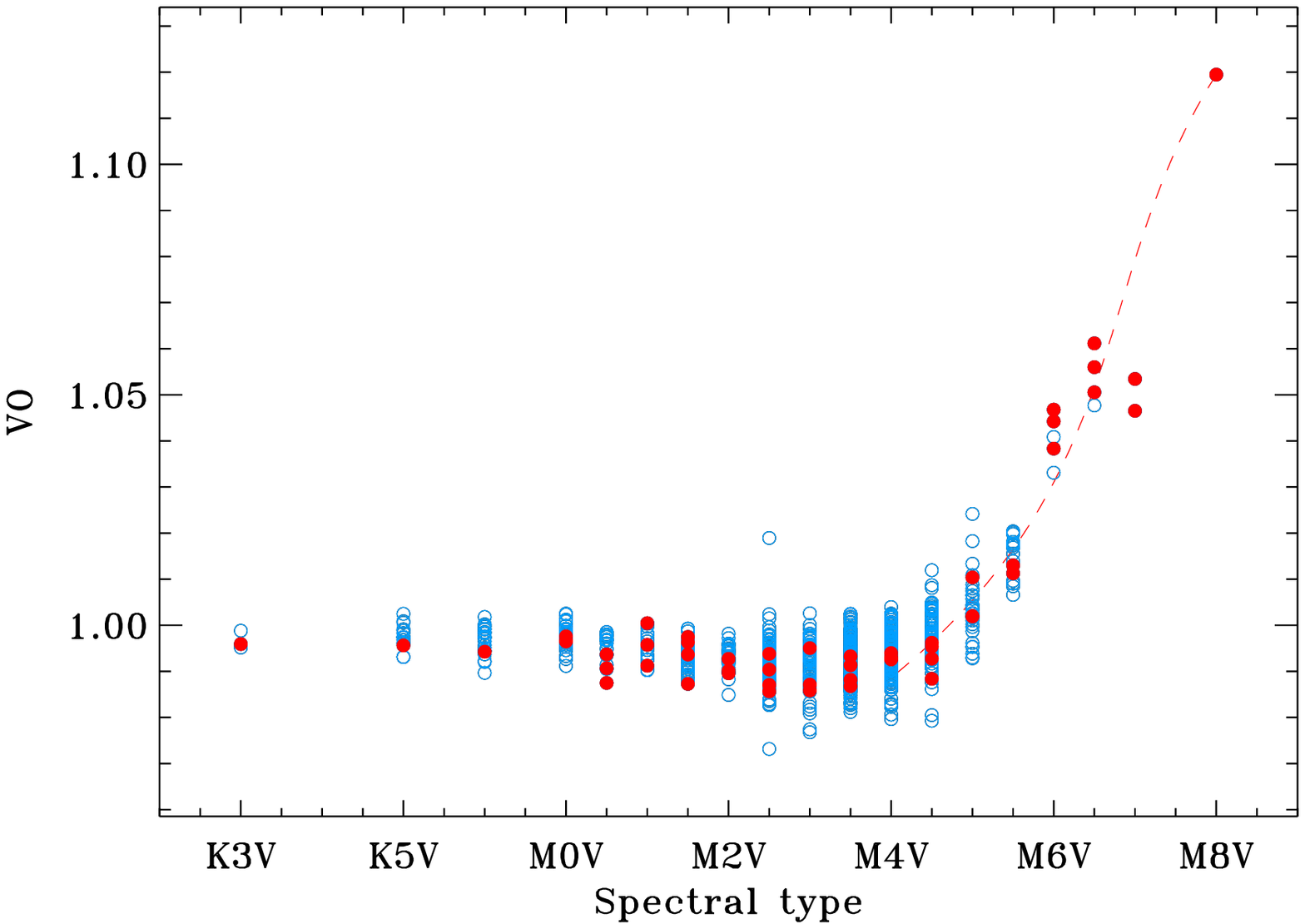}
\includegraphics[width=0.32\textwidth]{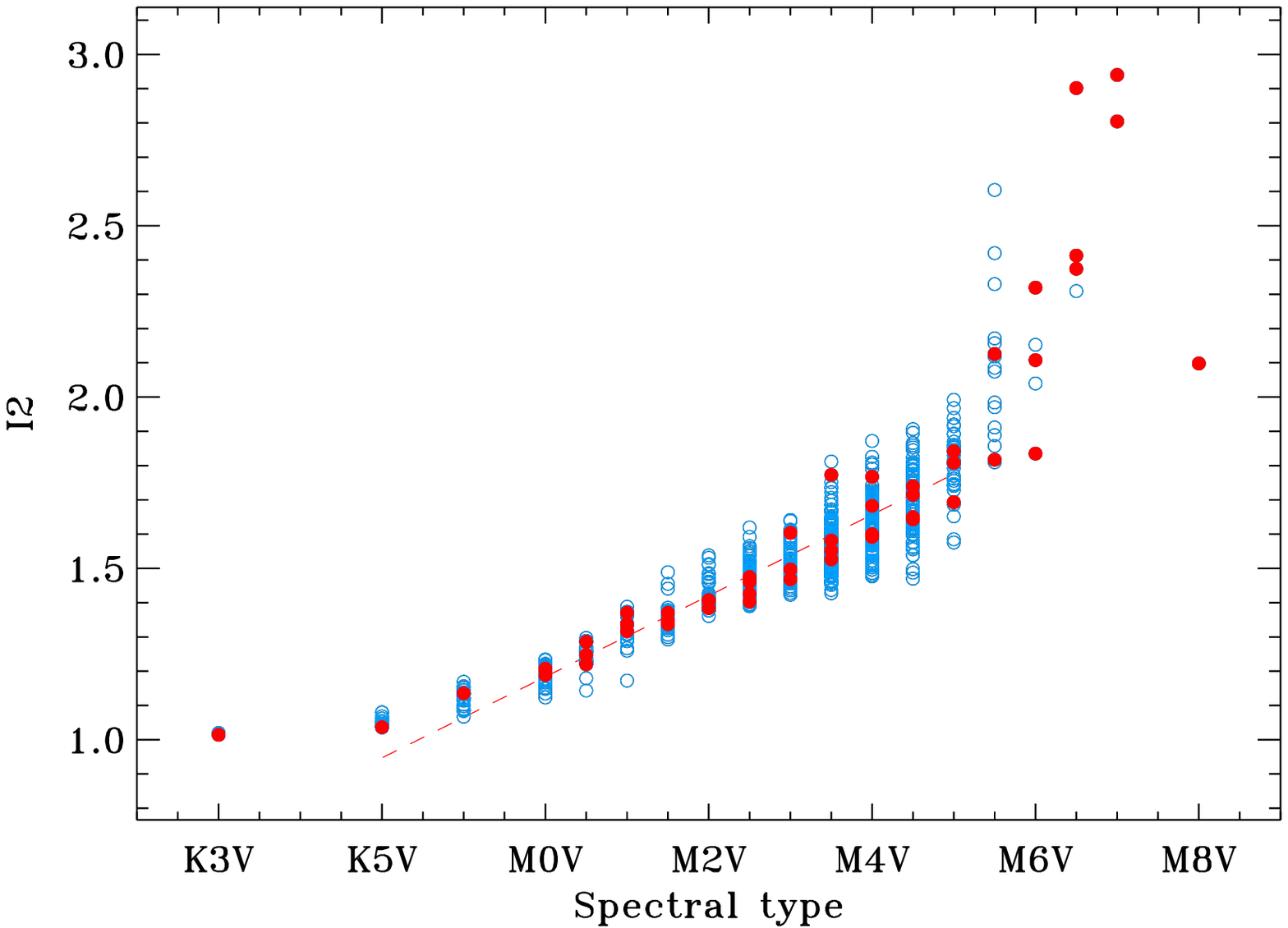}
\caption{\label{fig.indicesvsspt} Nine representative spectral indices as a function of spectral type.
Filled (red) circles: standard stars.
Open (blue) circles: remaining target stars.
Dashed (red) line: fit to straight line, parabola, or cubic polynomial, drawn only in the range of application of the fit.  
{\em Top left and middle:} TiO\,2 and TiO\,5 (Reid et~al. 1995), with negative slopes and useable up to M7\,V;
{\em top right:} PC1 (Mart\'{\i}n et~al. 1996), a monotonous spectral indicator from K5\,V to M8\,V; 
{\em centre left and middle:} VO-7912 (Mart\'{\i}n et~al. 1999) and Color-M (L\'epine et~al. 2003), two indices very similar to the PC1. Note the logarithmic scale in Color-M;
{\em centre right:} ratio B (Kirkpatrick et~al. 1991), sensitive to several stellar parameters and, thus, with a large scatter in the spectral type relation;
{\em bottom left:} CaH\,3 (Reid et~al. 1995), with a slightly larger scatter than pseudo-continuum or titanium oxide indices, due to metallicity;
{\em bottom middle:} VO (Kirkpatrick et~al. 1995), useable only for determining spectral types later than M4\,V;
{\em bottom right:} I2 (Mart\'{\i}n \& Kun 1996), with a linear range of variation from mid-K to mid-M and a sudden increase (or high dispersion) at late-M.
}
\end{figure*}

For every star observed with CAFOS, we computed the stellar numerator and denominator fluxes using an automatic trapezoidal integration procedure.
When all indices were available, we plotted all spectral index vs. spectral type diagrams for the standard stars listed in Table~\ref{table.standards} and fitted low-order polynomials to the data points.
Although some spectral indices allowed linear (e.g., I2, Ratio~B) or parabolic fits, most of the fits were to cubic polynomials of the form SpT($i$) = $a + b\,i + c\,i^2 + d\,i^3$, where $i$ was the index. 
In all cases, we checked our diagrams and fits with those in the original papers and found no significant differences (of less than 0.5 subtypes). 
We also took special care in defining the range of application of our fits in spectral type. 
The different shapes of fitting curves, ranges of application, and internal dispersion of the data points are illustrated in Fig.~\ref{fig.indicesvsspt}.

Some indices are sensitive not only to spectral type (i.e., effective temperature), but also to surface gravity (e.g., I2, Ratio~A, CaH\,Narr, Ratio~C, Na-8190 -- Sect.~\ref{subsec:gravity}), metallicity (e.g., CaOH, CaH\,1, CaH\,2, CaH\,3 -- Sect.~\ref{subsec:metallicity}), or activity (H$\alpha$ -- Sect.~\ref{subsec:activity}).
We identified the spectral indices with the widest range of application and least scatter. 
Table~\ref{table.fits} lists the coefficients of the cubic polynomial fits of the five spectral indices that we eventually chose for spectral typing (note the logarithmic scale of the Color-M index).
The spectral index vs. spectral type diagrams of VO-7912 and Color-M are very similar to those of PC, TiO\,2, and TiO\,5, shown in Fig.~\ref{fig.indicesvsspt}. 
All of them are valid from K7\,V to M7\,V (to M8\,V in the case of PC1 and Color-M), while the dispersion of the fits is of about 0.5 subtypes. 
The TiO\,5 has been a widely used index for spectral typing (Reid et~al. 1995; Gizis 1997; Seeliger et~al. 2011; L\'epine et~al. 2013) but, to our knowledge, we propose here for the first time to use it with a nonlinear fit.

In Table~\ref{table.indices}, we list the values of the five spectral-typing indices of all CAFOS stars together with the CaH\,2 and CaH\,3 indices that are used to compute the $\zeta$ metallicity index (Sect.~\ref{subsec:metallicity}) and the pseudo-equivalent width of the H$\alpha$ line (Sect.~\ref{subsec:activity}).

\begin{table}[]
        \centering
        \caption{Coefficients and standard deviation (in spectral subtypes) of the cubic polynomial fits of the five spectral-typing indices$^{a}$.}
         \label{table.fits} 
        \begin{tabular}{ l c c c c c}
         \hline
         \hline
         \noalign{\smallskip}
Index           & $a$           & $b$   & $c$   & $d$   & $\sigma$         \\ 
\noalign{\smallskip}            
          \hline
         \noalign{\smallskip}
TiO\,2          & +11.0         & --22  & +28   & --20  & 0.83          \\ 
TiO\,5          & +9.6          & --20  & +17.0   & --9.0   & 0.57          \\ 
PC1                     & --50          & +97   & --59  & +12.4 & 0.52            \\ 
VO-7912                 & --520                 & +1300         & --1070  & +300  & 0.59          \\ 
Color-M                 & +1.98                 & +13.1         & --15.6  & +10.3 & 0.53          \\ 
\noalign{\smallskip}
\hline
        \end{tabular}
\begin{list}{}{}
\item[$^{a}$] 
We used the relation SpT($i$) = $a + b\,i + c\,i^2 + d\,i^3$ for the cubic fits.
For the Color-M index we used the relation SpT($i$) = $a + b\,\log{i} + c\,(\log{i})^2 + d\,(\log{i})^3$. 
\end{list}
\end{table}

\subsubsection{Adopted spectral types}
\label{subsec:spt}

After applying the best-match and $\chi^{2}_{\rm min}$ methods and using the five spectral-index-type relations in Table~\ref{table.fits}, we obtained seven complementary spectral-type determinations for each star. 
In Table~\ref{table.SpT}, we assigned a value between 0.0 and 8.0 in steps of 0.5 to each M spectral subtype for all stars of dwarf luminosity class (there are no stars later than M8.0\,V in our CAFOS sample).
In addition, we used the values --2.0 and --1.0 for referring to K5\,V and K7\,V spectral types (there are no K6, 8, 9 spectral types in the standard K-dwarf classification -- Johnson \& Morgan 1953; Keenan \& McNeil 1989).
In some cases, we were able to identify dwarfs earlier than K5\,V with the best-match and $\chi^{2}_{\rm min}$ methods.
For giant stars, we only provide a visual estimation (K\,III, M\,III) based on the spectral types of well-known giant standard stars observed with CAFOS.

For all late-K and M dwarfs, we calculated one single spectral type per star based on the information provided by the seven individually determined spectral indices.
We used a median of the seven values for cases where they were identical within 0.5 subtypes.
To avoid any bias, we carefully checked the original spectra if the spectral types from the best-match and $\chi^{2}_{\rm min}$ methods and from the spectral indices were different by 0.5 subtypes, or if any spectral type deviated by 1.0 subtypes or more (which occurred very rarely).
In these cases, we adopted the spectral type of the closest (visually and in $\chi^2$) standard star. 
The uncertainty of the adopted spectral types is 0.5 subtype, except for some odd spectra indicated with a colon (probably young dwarfs of low gravity or subdwarfs of low metallicity; see below).


\section{Results and discussion}

\subsection{Spectral types}
\label{subsection:SpT}

\begin{figure}[]
\includegraphics[trim= 25 0 0 0,width=0.48\textwidth]{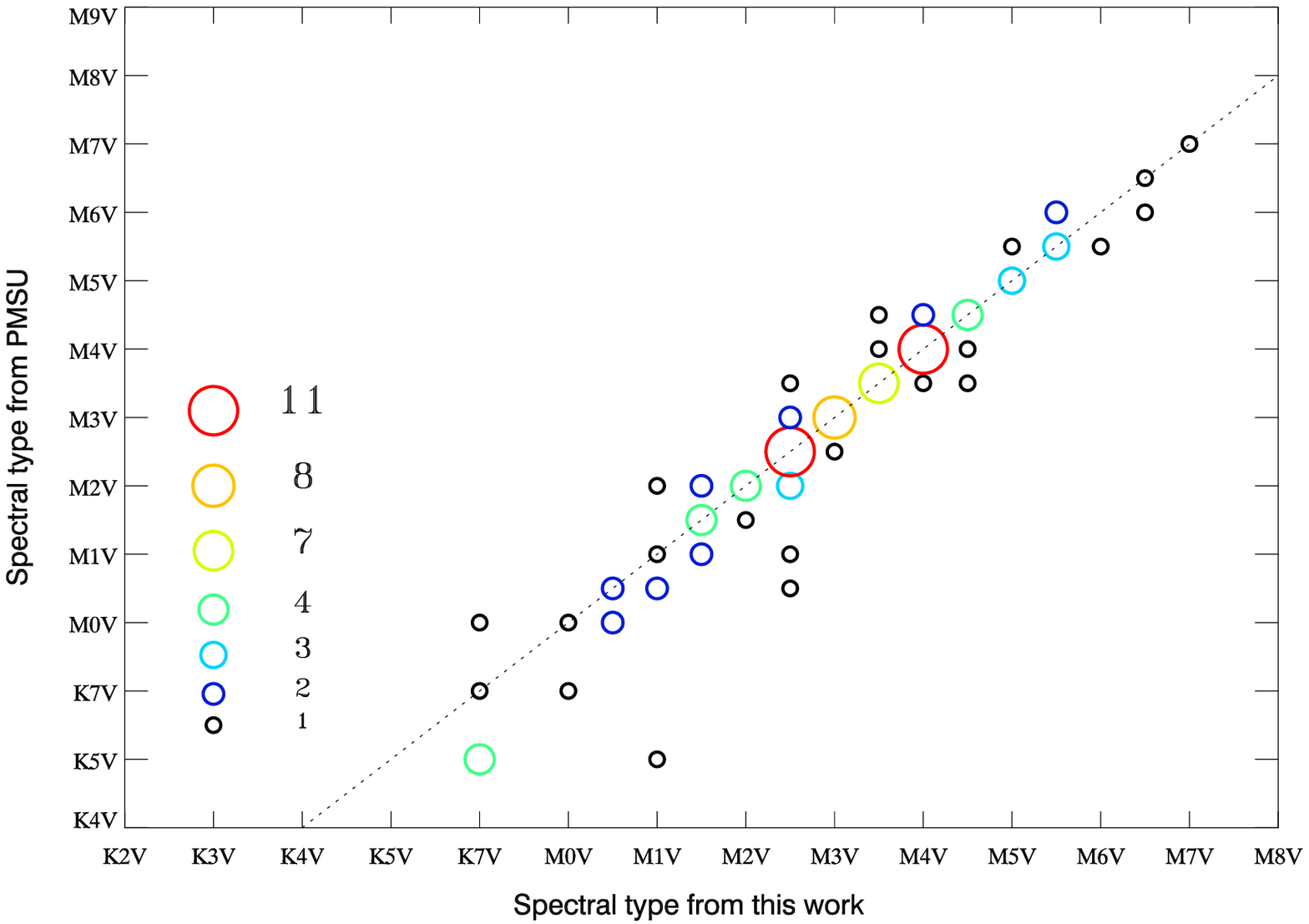}
\includegraphics[trim= 25 0 0 0,width=0.48\textwidth]{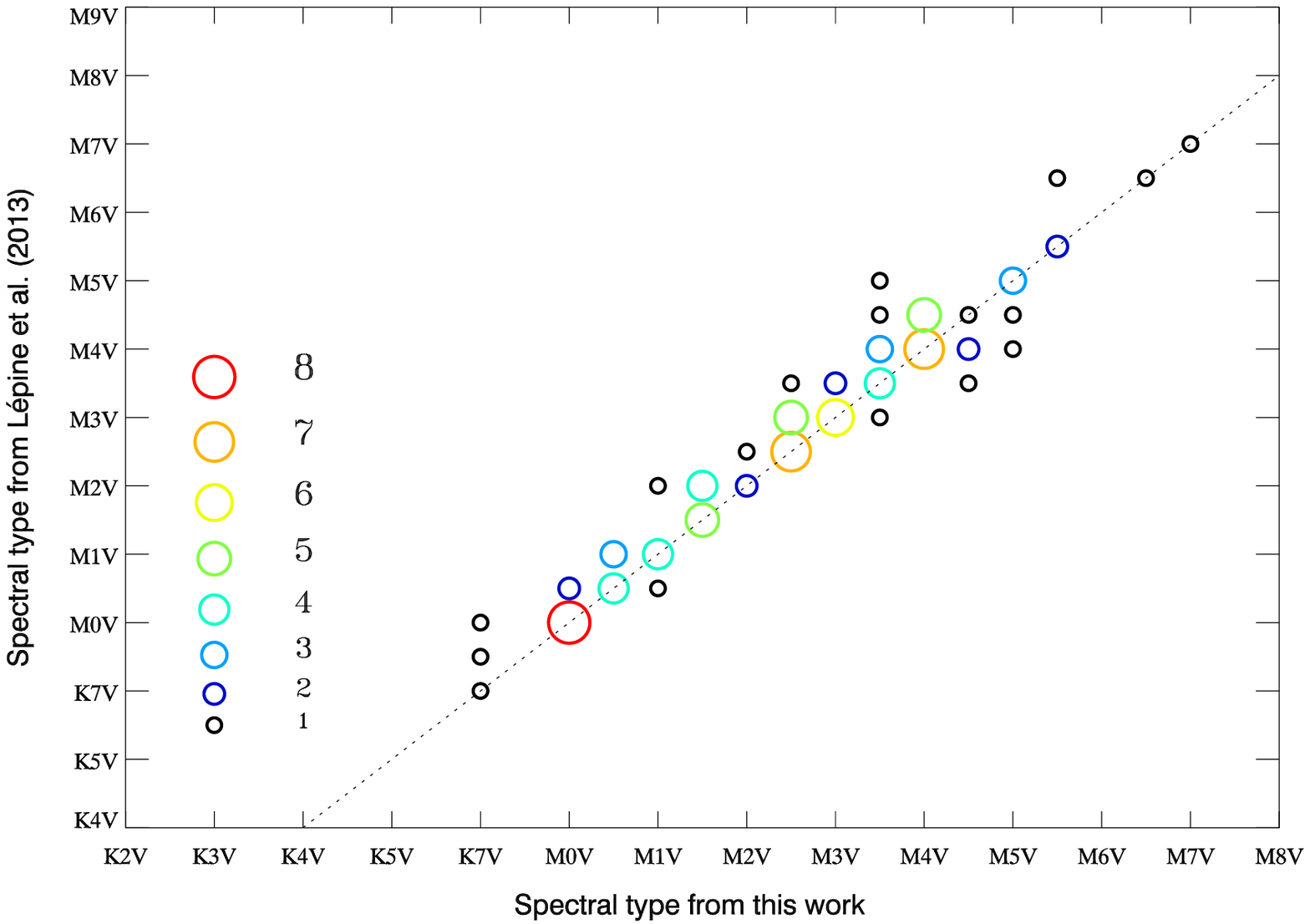}
\includegraphics[trim= 25 0 0 0,width=0.48\textwidth]{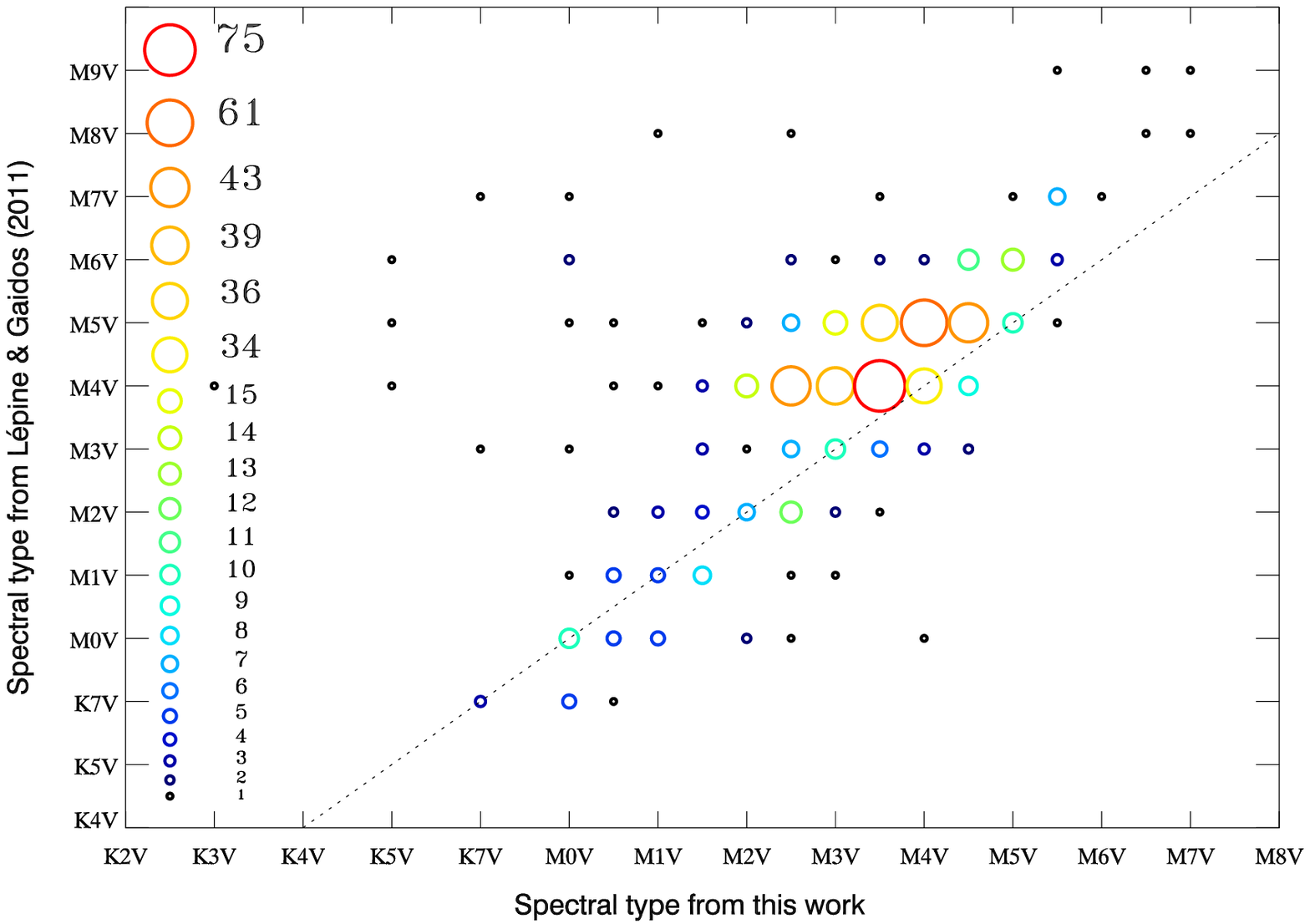}
\caption{\label{fig.SpTcomparison} Spectral type comparison between our results and those from PMSU ({\em top panel}), L\'epine et~al. (2013; {\em middle panel}), and L\'epine \& Gaidos (2011; from $V^*-J$ photometry, {\em bottom panel}).
The larger a circle, the greater the number of stars on a data point.
Dotted lines indicate the one-to-one relationship.} 
\end{figure}

Among the {753} investigated stars with adopted spectral types from CAFOS data, there were
{23} late-K dwarfs at the K/M boundary (K7\,V),
{21} early- and intermediate-K dwarfs (K0--5\,V),
{22} M-type giants (M\,III), 
{three} K-type giants (K\,III), and
{one} star without class determination (i.e., J04313+241\,AB).
This left {683} M-type dwarfs (and subdwarfs) in our CAFOS sample. 

As shown in Table~\ref{table.SpT}, we searched for previous spectral type determinations in the literature for the {753} investigated stars (small `m' and 'k' denote spectral types estimated from photometry).
Taking previous determinations and estimations into account, we derived spectral types from spectra for the first time for {305} stars, and revised typing for most of the remaining {448} stars.

The agreement in spectral typing with previous large spectroscopic surveys of M dwarfs is shown in Fig.~\ref{fig.SpTcomparison}.
The standard deviations of the differences between the spectral types derived by us and by PMSU (with a narrower wavelength interval; 100 stars in common) and by L\'epine et~al. (2013; 95 stars in common) were {0.55} and {0.38} subtypes, respectively, which are of the order of our internal uncertainty (0.5 subtypes).
The standard deviation of the differences between our spectral types and those estimated from photometry by L\'epine \& Gaidos (2011; 576 stars in common) is larger, of up to {1.32 subtypes}.
The bias towards later spectral types in L\'epine \& Gaidos (2011) and the scatter of the spectral type differences is obvious from the bottom panel in Fig.~\ref{fig.SpTcomparison}. 
In particular, we measured maximum differences of up to 7 subtypes, by which some late-M dwarf candidates become actual K dwarfs (probably due to the use by L\'epine \& Gaidos of $B_J$ and $R_F$ from photographic plates for the spectral type estimation; see references in Sect.~\ref{section.sample}).
However, over 93\,\% of the compared stars have disagreements lower than or equal to 2 subtypes.
We emphasize that our CARMENCITA data base is very homogeneous because more than 95\,\% of the spectral type determinations come from either PMSU, L\'epine et~al. (2013), or our CAFOS data, which are consistent with each other, as shown above.

Of the {683} CAFOS M-type dwarfs (and subdwarfs), {414} and {106} M dwarfs satisfy our criteria in Table 1 of restrictive $J$-band spectral type limiting and completeness, respectively. 
In total, {261} dwarfs have spectral type M4.0\,V or later. 
The brightest, latest of them are being followed-up with high-resolution spectrographs and imagers and with data from the bibliography
to identify the most suitable targets for CARMENES (no physical or visual companions at less than 5\,arcsec, low $v \sin{i}$; see forthcoming papers of this series).
Furthermore, there are {61} relatively bright ($J <$ 10.9\,mag) CAFOS stars with spectral types between M5.0\,V and M8.0\,V that are also suitable targets for any other near-infrared radial-velocity monitoring programmes with the instruments mentioned above (i.e., HPF, SPIRou, IRD).

\subsection{Gravity}
\label{subsec:gravity}

\begin{table}[]
        \centering
        \caption{Giant stars observed with CAFOS. }
         \label{table.giants} 
        \begin{tabular}{l l l l}
         \hline
         \hline
         \noalign{\smallskip}
Karmn           & Name                          & Giant                 \\ 
\noalign{\smallskip}            
          \hline
         \noalign{\smallskip}
J00146+202      & $\chi$ Peg                    & Standard              \\ 
J00367+444      & V428 And                      & Standard              \\ 
J00502+601      & HD 236547                     & Standard              \\ 
J01012+571      & 1RXS J010112.8+570839 & New           \\ 
J01097+356      & Mirach                                & Standard                 \\ 
J02479--124     & Z Eri                         & Standard              \\ 
J02558+183      & $\rho^{02}$ Ari               & Standard              \\ 
J03319+492      & TYC 3320--337--1              & LG11$^{a}$            \\ 
J04206--168     & DG Eri                                & Standard                 \\ 
J07420+142      & NZ Gem                        & Standard              \\ 
J10560+061      & 56 Leo                                & Standard                 \\ 
J11018--024     & $p^{02}$ Leo                  & Standard              \\ 
J11201+301      & HD 98500                      & Standard              \\ 
J11458+065      & $\nu$ Vir                     & Standard              \\ 
J12322+454      & BW CVn                        & Standard              \\ 
J12456+271      & HD 110964                     & Standard              \\ 
J12533+466      & BZ CVn                                & Standard                 \\ 
J13587+465      & HD 122132                     & Standard              \\ 
J17126--099     & Ruber 7                       & JE12$^{b}$             \\ 
J17216--171     & TYC 6238--480--1      & JE12$^{b}$            \\ 
J18423--013     & Ruber 8                               & JE12$^{b}$             \\ 
J22386+567      & V416 Lac                      & Standard              \\ 
J23070+094      & 55 Peg                                & Standard                 \\ 
J23177+490      & 8 And                         & Standard              \\ 
J23266+453      & 2MASS J23263798+4521054 & Background \\ 
\noalign{\smallskip}
\hline
        \end{tabular}
\begin{list}{}{}
\item[$^{a}$] LG11: L\'epine \& Gaidos (2011).
\item[$^{b}$] JE12: Jim\'enez-Esteban et~al. (2012).
\end{list}
\end{table}

\begin{figure}
\includegraphics[trim= 25 0 0 0,width=0.5\textwidth]{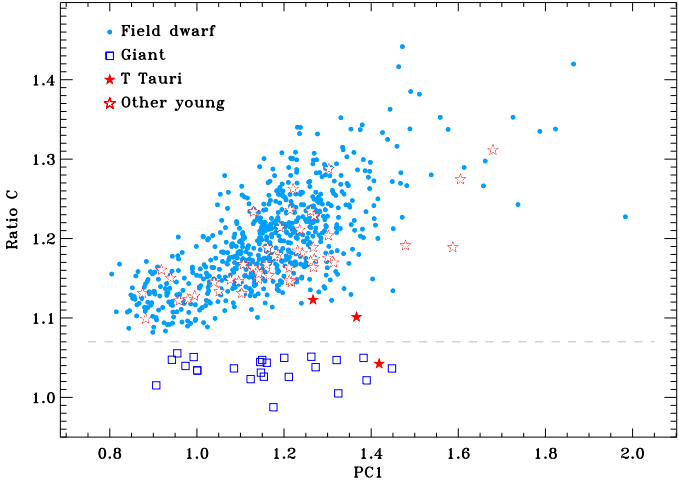}
\caption{\label{fig.RatioCvsPC1} Ratio~C vs. PC1 index-index diagram. 
The different symbols represent 
field dwarfs (small dots, blue), 
giants (open squares, dark blue),
Taurus stars (filled stars, red), and
other young stars (open stars, red).
All giants are below the dashed line at Ratio~C = 1.07. 
The dashed line is the empirical border of the giant star region.}
\end{figure}

Table~\ref{table.giants} lists the 25 giants observed with CAFOS. 
Of these, 17 stars have previously  been tabulated as M giant standard stars (e.g., Keenan \& McNeil 1989; Garc\'{\i}a 1989; Kirkpatrick et~al. 1991; S\'anchez-Bl\'azquez et~al. 2006).
They are bright ($J \lesssim$ 5.0\,mag; down to --1.0\,mag in the case of Mirach, $\beta$~And) and show the low-gravity spectral features typically found in M giants:
faint alkali lines (K~{\sc i} $\lambda$$\lambda$7665,7699\,{\AA} and Na~{\sc i} $\lambda$$\lambda$8183,8195\,{\AA}), a tooth-shaped feature produced by MgH/TiO blend near 4770\,{\AA}, and a decrease of CaH in the A-band at 6908--6946\,{\AA} with the increase of luminosity (Kirkpatrick et~al. 1991; Mart\'in et~al. 1999; Riddick et~al. 2007; Gray \& Corbally 2009).
Two other stars, V428~And and HD~236547, are well-known K giant standard stars (Jacoby et~al. 1984; Garc\'{\i}a 1989; Kirkpatrick et~al. 1991).

Of the other six giant stars in Table~\ref{table.giants}, three have $J$-band magnitudes of 6.1--6.2\,mag and were identified by Jim\'enez-Esteban et~al. (2012) as some of the reddest Tycho-2 stars with proper motions $\mu >$ 50\,mas\,a$^{-1}$, namely Ruber~7, TYC 6238--480--1, and Ruber~8 (which seems to be also one of the brightest metal-poor M giants ever identified).
The remaining three giant stars, with faint $J$-band magnitudes between 8.2 and 10.0\,mag, are listed below.
\begin{itemize}
\item J01012+571 (1RXS J010112.8+570839).
It is a previously unknown distant M giant close to the Galactic plane ($b$ = --5.7\,deg).
It was serendipitously identified in an unpublished photometric survey by one of us (J.A.C.), and was observed with CAFOS because of its very red optical and near-infrared colours and possible association with an X-ray event catalogued by {\em ROSAT} at a separation of only 6.4\,arcsec.
\item J03319+492 (TYC 3320--337--1). 
From photographic magnitudes, Pickles \& Depagne (2010) and L\'epine \& Gaidos (2011) classified it as an M1.9 and M3 dwarf, respectively.
However, it appears to be an early-K giant with a significant proper motion of 56\,mas\,a$^{-1}$.
It is not possible to separate it from the main sequence in a reduced proper-motion diagram.
\item J23266+453 (2MASS J23263798+4521054). 
Our intention was to observe BD+44\,4419\,B (G~216--43), an M4.5 dwarf of roughly the same $V$-band magnitude (10.3 vs. 10.9\,mag).  
Unfortunately, we incorrectly observed instead a background giant at a separation of about 20\,arcsec.
\end{itemize}

In a Ratio~C vs. PC1 index-index diagram as the one shown in Fig.~\ref{fig.RatioCvsPC1}, where Ratio~C is highly sensitive to gravity and PC1 is an effective temperature proxy (PC1 was indeed one of the five indices used for deriving spectral types), all giants are below the dashed line at Ratio~C = 1.07. 
There is only one star not classified as a giant that lies below that empirical boundary.
It is J04313+241 (V927 Tau AB), a T~Tauri star for which we did not provide a luminosity class in Sect.~\ref{subsection:SpT}.
We discuss this in detail in Sect.~\ref{subsec:activity}. Ratio~C, which contains the sodium doublet at 8193,8195\,{\AA}, can also be used as a youth indicator (e.g., Schlieder et~al. 2012b).

\subsection{Metallicity}
\label{subsec:metallicity}

In F-, G-, and K-type stars whose photospheric continua are well-defined in high-resolution spectra, stellar metallicity is computed through spectral synthesis (McWilliam 1990; Valenti \& Piskunov 1996; Gonz{\'a}lez 1997; Gonz\'alez Hern\'andez et~al. 2004; Valenti \& Fischer 2005; Recio-Blanco et~al. 2006) or measuring equivalent widths, especially of iron lines (Sousa et~al. 2008, 2011; Magrini et~al. 2010; Adibekyan et~al. 2012; Tabernero et~al. 2012; Bensby et~al. 2014).
However, it is not possible to measure a photospheric continuum in M-type stars and, thus, their metallicity is studied through other techniques.
Since the first determinations from broad-band photometry by Stauffer \& Hartmann (1986), there have been three main observational techniques employed to determine metallicity in M dwarfs: 
\begin{itemize}
\item Photometry calibrated with M dwarfs in physical double and multiple systems with warmer companions, typically F, G, K dwarfs, of known metallicity 
(Bonfils et~al. 2005;
Casagrande et~al. 2008;
Schlaufman \& Laughlin 2010;
Neves et~al. 2012).
\item Low-resolution spectroscopy, also calibrated with M dwarfs with earlier primaries, 
in the optical 
(Dhital et~al. 2012), 
in the near infrared 
(Rojas-Ayala et~al. 2010, 2012;
Terrien et~al. 2012;
Mann et~al. 2014;
Newton et~al. 2014),
or in both wavelength ranges 
(Mann et~al. 2013, 2015).
\item High-resolution spectroscopy in the optical 
(from spectral synthesis: Woolf \& Wallerstein 2005;
from spectral indices: Woolf \& Wallerstein 2006 and Bean et~al. 2006;
from the measurement of pseudo-equivalent widths: Neves et~al. 2013, 2014), 
in the near-infrared 
(Shulyak et~al. 2011 in the $Y$ band;
\"Onehag et~al. 2012 in the $J$ band;
Tsuji \& Nakajima 2014 in the $K$ band), 
or in the optical and near-infrared simultaneously
(Gaidos \& Mann 2014). 
The novel mid-resolution spectroscopy study in the optical aided with spectral synthesis by Zboril \& Byrne (1998) also belongs in this item. 
\end{itemize}

\begin{figure}
\includegraphics[trim= 25 0 0 0,width=0.5\textwidth]{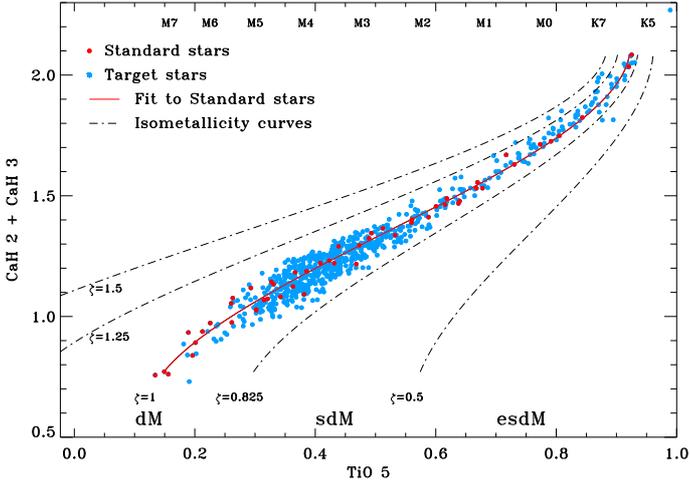}
\caption{\label{fig.zeta} CaH\,2\,+\,CaH\,3 vs. TIO\,5 index-index diagram of our CAFOS stars, after discarding giants.
The spectral types at the top are indicative, but follow the TiO\,5 fit given in Table~\ref{table.fits}. 
The solid and dash-dotted lines are iso-metallicity curves of the $\zeta$ index. }
\end{figure}

For the {753} CAFOS stars, we computed the $\zeta\rm_{TIO/CaH}$ metallicity parameter (denoted $\zeta$ for short) described by L\'epine et~al. (2007):
\begin{equation}
\label{eq.zeta}
\zeta = \frac{1-{\rm TIO\,5}}{1-[{\rm TiO\,5}]_{Z_\odot}},
\end{equation}
\noindent where [TiO\,5]$_{Z_\odot}$ = $0.571-1.697{\rm CaH}+1.841{\rm CaH}^2 -0.454{\rm CaH}^3$ is a third-order fit of CaH = CaH\,2\,+\,CaH\,3 for {\em our} standard stars, and TiO\,5, CaH\,2, and CaH\,3 are the spectral indices of Reid et~al. (1995) (see also Gizis \& Reid 1997 and L\'epine et~al. 2003).
The $\zeta$ index is correlated with metallicity in metal-poor M subdwarfs (Woolf et~al. 2009) and metal-rich dwarfs (L\'epine et~al. 2007; Mann et~al. 2014). 
For completeness, we also tabulate the $\zeta$ index for our 25 giants, but they are not useful for a comparison.
We made the same assumption of standard stars having solar metallicity ($\zeta \approx$ 1) as in L\'epine et~al. (2007), which was later justified by the small dispersion of the data points.

We looked for M dwarfs (and subdwarfs) in our sample with abnormal metallicity, which could be spotted in a CaH\,2\,+\,CaH\,3 vs. TIO\,5 index-index diagram as in Fig.~\ref{fig.zeta}.
L\'epine et~al. (2007) defined the classes subdwarf (sd), 0.5 $< \zeta <$ 0.825, and extreme subdwarf (esd), $\zeta <$ 0.5.
All our non-giant stars except two have $\zeta$ values greater than 0.825, which is the empirical boundary between dwarfs and subdwarfs.
The spectra of the two exceptions show shallower molecular bands and lines than M dwarfs of the same spectral type.

One of our two subdwarf candidates is J19346+045 (sdM1:, $\zeta$ = 0.775 -- HD\,184489). 
Some authors have reported features of low metallicity (e.g., Maldonado et~al. 2010), but none had classified it as a subdwarf (but see Sandage \& Kowal 1986).
Its low effective temperature has prevented spectral synthesis analyses on high-resolution spectra.

The other new subdwarf candidate is J16354--039 (sdM0:, $\zeta$ = 0.664 -- \object{HD\,149414\,B}, BD--03~3968B). 
Giclas et~al. (1959) discovered it and associated it with the G5\,Ve single-line spectroscopic binary HD\,149414\,Aa,Ab.
Afterwards, its membership in the very wide system has been investigated by Poveda et~al. (1994), Tokovinin (2008), and Dhital et~al. (2010), for example, and confirmed and quantified by Caballero (2009).
The projected physical separation between Aa,Ab and B amounts to 53\,000\,au (about a quarter of a parsec).
Remarkably, the primary is a halo binary of low metallicity ([Fe/H] $\sim$ --1.4 -- Strom \& Strom 1967; Sandage 1969; Cayrel de Strobel et~al. 1997; Holmberg et~al. 2009).
This  explains the low $\zeta$ metallicity index of J16354--039 for its spectral type and the wide separation of the system (due to gravitational disruption by the Galactic gravitational potential or to common origin and ejection from the same cluster; cf. Caballero 2009 and references therein).  

In addition, J12025+084 (M1.5\,V; $\zeta$ = 0.898 -- \object{LHS~320}) was classified by Gizis (1997) as an sdM2.0 star and was investigated extensively afterwards with high-resolution imagers (Gizis \& Reid 2000; Riaz et~al. 2008; Jao et~al 2009; Lodieu et~al. 2009). 
However, we failed to detect any subdwarf signpost in our high signal-to-noise spectrum, which is partly consistent with the metallicity [Fe/H] = --0.6$\pm$0.3 measured by Rajpurohit et~al. (2014).

No CAFOS star showed a very high metallicity index greater than $\zeta =$ 1.5.
In spite of the dispersion of the $\zeta$ index around unity, we considered that all our {726} dwarfs ({753} stars in total minus the 25 giants and the two subdwarfs) {\em approximately} have solar metallicity ([Fe/H] $\approx$ 0.0).
This assumption is relevant for instance to derive the mass from absolute magnitudes, the spectral types, and theoretical models that need metallicity as an input.

\subsection{Activity}
\label{subsec:activity}

Chromospheric activity is one of the main relevant parameters for exoplanet detection around M dwarfs.
The heterogeneities on the stellar surface of the almost-fully convective, rotating, M dwarfs, such as dark spots, may induce spurious radial-velocity variations at visible wavelengths (Bonfils et~al. 2007; Reiners et~al. 2010; Barnes et~al. 2011; Andersen\& Korhonen 2015; Robertson et~al. 2015).
Near-infrared observations are expected to improve the precision of radial-velocity measurements with respect to the visible for stars cooler than M3, and CARMENES will cover the wavelength range from 0.55 to 1.70\,$\mu$m.
In spite of this, we plan to identify the least active stars for our exoplanet search.
Moreover, several authors have identified significant differences between colours and spectral indices of active and inactive stars of similar properties that may affect the spectral typing of M dwarfs (Stauffer \& Hartmann 1986; Hawley et al. 1996; Bochanski et~al. 2007; Morales et al. 2008).

\subsubsection{H$\alpha$ emission}

\begin{figure}
\includegraphics[trim= 25 0 0 0,width=0.5\textwidth]{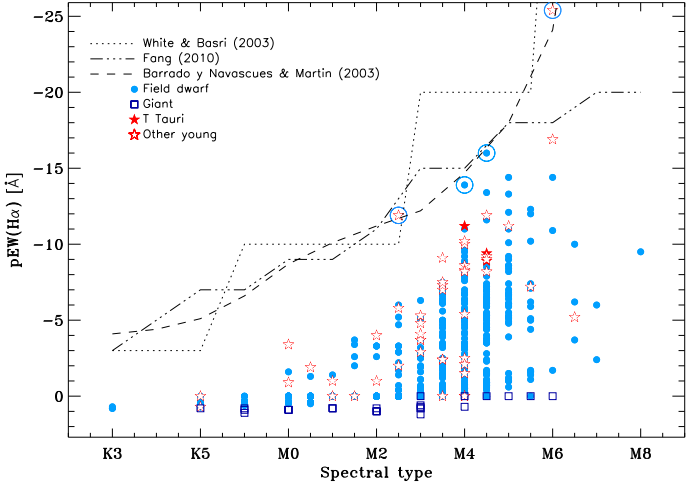}
\caption{\label{fig.pewHalpha} Pseudo-equivalent width of H$\alpha$ line vs. spectral type diagram.
Measurement errors are 0.5 subtypes for the spectral type and are of the order of 10\,\% with a minimum of 0.1\,{\AA} for $pEW$(H$\alpha$) (but scatter due to variability is probably larger than 10\,\%).
Dotted, dash-dotted, and dashed lines indicate the boundaries between chromospheric and accretion emission for different authors.
Giants, plotted with open squares, have filled H$\alpha$ lines or in absorption.
None of the young stars, plotted with open and filled stars, show accretion emission.
The four stars at the accretion boundary discussed in the text are encircled.}
\end{figure}

The H$\alpha$ index (Reid et~al. 1995) was one of the 31 indices measured on our CAFOS spectra (Table~\ref{table.index}).
For a better reliability on the activity determination and comparison with other works, we also measured the pseudo-equivalent width of the H$\alpha$ line, $pEW$(H$\alpha$), of all the CAFOS stars (Table~\ref{table.indices}).
We here used $pEW$(H$\alpha$) as the proxy for activity (we used the pseudo-equivalent width of the line measured with respect to a local pseudo-continuum instead of the equivalent width because in M --and L, T, and Y-- dwarfs the spectral continuum is not observable -- e.g. Tsuji \& Nakajima 2014).

We plot in Fig.~\ref{fig.pewHalpha} the $pEW$(H$\alpha$) vs. spectral type diagram for the whole sample.
M dwarfs with late spectral types tend to show the H$\alpha$ line in (strong) emission more often than earlier stars (see e.g. Hawley et~al. 1996, West et~al. 2004, or Reiners et~al. 2013).
There is a significant number of M4.0\,V stars and later,
however, that show very low H$\alpha$ emission below 5\,{\AA} (in absolute values).  

A few stars stand out in the $pEW$(H$\alpha$) vs. spectral type diagram in Fig.~\ref{fig.pewHalpha}. 
Four of them lie at the boundary between chromospheric and accretion emission, as defined by Barrado y Navascu\'es \& Mart\'in (2003), White \& Basri (2003), and Fang (2010). 
Their high activity led us to investigate them in detail.
\begin{itemize}
\item J04290+186 (M2.5\,V, $pEW$(H$\alpha$) = --11.9$^{+0.5}_{-0.3}$\,{\AA}, \object{V1103~Tau}).
It is a  member of the 600\,Ma-old Hyades cluster (Johnson et~al. 1962; Griffin et~al. 1988; Stauffer et~al. 1991; Reid 1992).
\item J04544+650 (M4.0\,V, $pEW$(H$\alpha$) = --13.9$^{+0.8}_{-0.5}$\,{\AA}, \object{1RXS~J045430.9+650451}).
It is an anonymous Tycho-2 star (TYC 4087--1172--1; L\'epine \& Gaidos 2011) that we cross-matched with an aperiodic, variable, X-ray source identified by Fuhrmeister \& Schmitt (2003).
This X-ray variability and the presence of He~{\sc i} $\lambda$5875.6\,{\AA} in emission indicates that J04544+650 was flaring during our observations. 
\item J01567+305 (M4.5\,V, $pEW$(H$\alpha$) = --16.0$\pm$0.4\,{\AA}, NLTT~6496, \object{Koenigstuhl 4~A}).
It forms a loosely bound common-proper-motion pair together with the M6.5:\,V star NLTT~6491 (Koenigstuhl 4~B), and is associated with an X-ray source (Caballero 2012).
Interestingly, Aberasturi et~al. (2014) collected low-resolution spectroscopy for J01567+305 just two months
earlier, for which they determined a spectral type identical to ours within the uncertainties, but measured $pEW$(H$\alpha$) = --9.3$\pm$0.3\,{\AA}, which is significantly lower than our measurement. 
Our CAFOS spectrum also shows He~{\sc i} in emission, so the mid-M dwarf likely underwent a flare during our observations.
\item J07523+162 (M6.0\,V, $pEW$(H$\alpha$) = --25.4$^{+1.4}_{-1.0}$\,{\AA}, \object{LP~423--031}).
It has also been classified as a single M7\,Ve star from optical spectra (Cruz et~al. 2003; Reid et~al. 2003; Gatewood \& Coban 2009; Reiners \& Basri 2009), but as an M6\,V with surface gravity consistent with normal field dwarfs from near-infrared spectra (Allers \& Liu 2013).
From high-resolution spectroscopy ($pEW$(H$\alpha$) = --22.3\,{\AA}) and {\em ROSAT} X-ray count rates, Shkolnik et~al. (2009) assigned J07523+162 an age of about 100\,Ma, younger than the Pleiades.
However, Reiners \& Basri (2010) observed flaring activity in a J07523+162 spectrum ($pEW$(H$\alpha$) = --44.4\,{\AA}) and Gagn\'e et~al. (2014) and Klutsch et~al. (2014) were not able to determine membership in any known stellar kinematic group.
\end{itemize}
There is an additional fifth active dwarf that stands out among the remaining stars in Fig.~\ref{fig.pewHalpha}. 
It is J03332+462 (M0.0\,V, $pEW$(H$\alpha$) = --3.4$^{+0.5}_{-0.3}$\,{\AA}, \object{V577~Per\,B}), a confirmed member of the $\sim$70\,Ma-old AB~Doradus moving group (Zuckerman et~al. 2004; da~Silva et~al. 2009; Schlieder et~al. 2012a).
Its relatively bright primary at about 9\,arcsec is a young K2\,V star with strong ultraviolet and X-ray emission and lithium in absorption (Pounds et~al. 1993; Jeffries 1995; Montes et~al. 2001; Zuckerman \& Song 2004; Xing \& Xing 2012).

\subsubsection{Young (and very young) stars}

\begin{table}[]
        \centering
        \caption{Reported young stars in our sample. }
         \label{table.youngstars} 
        \begin{tabular}{l l l l}
         \hline
         \hline
         \noalign{\smallskip}
Karmn                   & Moving group /                & Ref.$^{a}$         \\
                                & association /  cluster /      &                         \\
                                & star-forming region   &                       \\
\noalign{\smallskip}            
          \hline
         \noalign{\smallskip}
J03332+462              & {AB Dor MG}                   & See text\\ 
J03466+243\,AB  & Pleiades                      & vMa45         \\ 
J03473--019             & AB Dor MG                     & Zuc04 \\ 
J03548+163\,AB  & {Hyades}                      & Gic62 \\ 
J04123+162\,AB  & Hyades                                & Gic62 \\ 
J04177+136\,AB  & {Hyades}                      & Gic62 \\ 
J04206+272              & Taurus                                & Sce07   \\ 
J04207+152\,AB  & {Hyades}                      & Gic62 \\ 
J04227+205              & {Hyades}                      & Reid93        \\ 
J04238+149\,AB  & Hyades                                & Gic62 \\ 
J04238+092\,AB  & {Hyades}                      & Gic62 \\ 
J04252+172\,ABC         & {Hyades}                      & Gic62 \\ 
J04290+186              & {Hyades}                      & Gic62 \\ 
J04313+241\,AB  & Taurus                                & HR72  \\ 
J04360+188              & {Hyades}                      & Pels75        \\ 
J04366+186              & {Hyades}                      & See text \\ 
J04373+193              & {Hyades}                      & Reid93        \\ 
J04393+335              & Taurus                                & Wic96   \\ 
J04425+204\,AB  & {Hyades}                      & Reid93        \\ 
J04430+187\,AB  & {Hyades}                      & Gic62 \\ 
J05019+011              & $\beta$~Pic MG                & Sch12 \\ 
J05062+046              & $\beta$~Pic MG                & Sch12 \\ 
J05256--091\,AB         & AB Dor MG                     & Shk12 \\ 
J05320--030             & $\beta$~Pic MG                & daS09 \\ 
J05415+534              & Her-Lyr MG?                   & Eis13 \\ 
J05457--223             & UMa MG                        & Tab15 \\ 
J06075+472              & AB Dor MG                     & Sch12         \\ 
J06246+234              & Young                         & Mon01 \\ 
J07319+362N             & Castor MG                     & Cab10 \\ 
J07319+362S\,AB & Castor MG                     & Cab10 \\ 
J07361--031             & Castor MG                     & Cab10 \\ 
J07523+162              & {Young}                               & See text        \\ 
J08298+267              & Castor MG                     & Cab10 \\ 
J09328+269              & Her-Lyr MG                    & Eis13 \\ 
J09362+375              & Young                         & Malo14        \\ 
J10196+198\,AB  & Castor MG                     & Cab10 \\ 
J10359+288              & $\beta$~Pic MG                & Sch12 \\ 
{J10508+068}            & Her-Lyr MG?                   & Eis13 \\ 
J11046--042S\,AB        & Her-Lyr MG                    & Eis13 \\ 
J13143+133\,AB  & Young                         & Sch14 \\ 
J15079+762              & IC~2391 MG                    & Mon01 \\ 
J17198+265              & Hyades        SC                      & Klu14   \\ 
J17199+265              & Hyades        SC                      & Klu14   \\ 
J18313+649              & AB Dor                                & Sch12   \\ 
J21376+016              & $\beta$~Pic MG                & Sch12 \\ 
J22160+546              & Her-Lyr MG?                   & Eis13 \\ 
J22234+324\,AB  & AB Dor MG                     & Malo14        \\ 
J23194+790              & Carina/Columba Ass.   & Klu   \\ 
J23209--017\,AB         & Argus Ass.                    & Malo14        \\ 
J23228+787              & Carina/Columba Ass.   & Klu   \\ 
\noalign{\smallskip}
\hline
        \end{tabular}
\begin{list}{}{}
\item[$^{a}$] {References} -- 
vMa45: van~Maanen 1945;
Gic62: Giclas et~al. 1962;
HR72: Herbig \& Rao 1972;
Pels75: Pels et~al. 1975;
Reid93: Reid 1993;
Wic96: Wichmann et~al. 1996; 
Mon01: Montes et~al. 2001;
Zuc04: Zuckerman et~al. 2004;
Sce07: Scelsi et~al. 2007;
daS09: da~Silva et~al. 2009;
Cab10: Caballero 2010;
Sch12: Schlieder et~al. 2012a;
Shk12: Shkolnik et~al. 2012;
Eis13: Eisenbeiss et~al. 2013;
Klu14: Klutsch et~al. 2014;
Malo14: Malo et~al. 2014;
Sch14: Schlieder et~al. 2014;
Tab14: Tabernero et~al. 2015;
Klu: Klutsch, priv. comm.
Part of the content of this table was extracted from Hidalgo (2014).
\end{list}
\end{table}

The identification of {one} open cluster member, {one} moving group member, and {one} purported young star in the field among five M dwarfs led us to examine the bibliography for other young star candidates in our CAFOS sample.
The result of this bibliographic search is summarised in Table~\ref{table.youngstars}. 
In total, {49} spectroscopically investigated stars in this work have been reported to belong to the 
Taurus-Auriga star-forming region ($\sim$1--10\,Ma, {three} stars),
$\beta$~Pictoris moving group ($\sim$12--22\,Ma, {five} stars),
Carina or Columba associations ($\sim$15--50\,Ma, {two} stars), 
Argus association ($\sim$40\,Ma, {one} star),
AB~Doradus moving group ($\sim$70--120\,Ma, {five} stars),
Pleiades cluster ($\sim$120\,Ma, {one} star -- with a relatively early K5\,V spectral type),
IC~2391 supercluster ($\sim$100--200\,Ma, {one} star),
Hercules-Lyra moving group ($\sim$200--300\,Ma, {four} stars -- note the question marks),
Castor moving group ($\sim$200--300\,Ma, {six} stars),
Ursa Major moving group ($\sim$300--500\,Ma, {one} star), and
Hyades cluster and supercluster ($\sim$600\,Ma, {14} and {2} stars, respectively),
and {four} to the young ($\tau \lesssim$ 600\,Ma) field star population in the solar neighbourhood.
See Zuckerman \& Song (2004) and Torres et al~al. (2008) for reviews on young moving groups.

The actual existence of some of the entities above (e.g., Hercules-Lyra and Castor moving groups, and IC~2391 and Hyades superclusters) is questioned by several authors.
Three of the five stars with the lowest H$\alpha$ emission for their spectral type belong to the hypothetical Castor moving group (Barrado y Navascu\'es 1998; Montes et~al. 2001; Ribas 2003; Caballero 2010; Mamajek 2013; Zuckerman et~al. 2013), and the other two to the Hyades (super-) cluster (van~Altena 1966; Hanson 1975; Legget \& Hawkins 1988; Hawley et~al. 1996; Stauffer et~al. 1997; Montes et~al. 2001; Klutsch et~al. 2014).
However, the extreme youth of some targets is confirmed by detection of lithium in absorption, X-ray in emission, and common proper-motion to bona~fide primaries in nearby young moving groups. 

Eleven of the 16 Hyads are known to be binaries.
The relatively large number and (apparent) high binary frequency is a natural consequence of the Malmquist bias, which leads to the preferential detection of intrinsically bright objects.
Equal-mass binaries are brighter than single stars of the same spectral type (by up to 0.75\,mag) and, thus, the frequency of binarity in our magnitude-limited sample is higher than in a bias-free, volume-limited sample.
While most of our targets lie at 20--30\,pc (Cort\'es-Contreras et~al. 2015), the overbrightness of binary Hyads makes them to look as if they were located roughly at 30\,pc instead at the nominal distance of the Hyades at about 46\,pc.
We suggest to investigate the actual multiplicity status of the five remaining {\em single} M dwarfs with a mid-resolution spectroscopic monitoring.

At $d \sim$ 140\,pc, the three Taurus stars in Table~\ref{table.youngstars} are {\em not} in the solar neighbourhood.
Since they are still on the Hayashi track of contraction, their radii are larger than those of dwarfs of the same effective temperature.
As a result, they are also much more luminous, which explains why we were able to observe them even though they are located an order of magnitude farther away than the rest of our dwarf targets. 
As expected from their extreme youth, the three T~Tauri stars have H$\alpha$ emissions in the highest quartile ($pEW$(H$\alpha$)s between --9 and --11\,{\AA}, and spectral types between M4.0 and M4.5) and have been investigated spectroscopically earlier (Herbig \& Rao 1972; Mathieu 1994; Wichmann et~al. 1996; Kenyon et~al. 1998; Scelsi et~al. 2008; Sestito et~al. 2008).
The three of them displayed not only H$\alpha$ in emission, but also H$\beta$ and H$\gamma$ (we used one of them, J04393+335, in Fig.~\ref{fig.bestmatch} to illustrate best the discarded wavelength ranges that are contaminated by activity in Sect.~\ref{subsec:least_square_minimisation}).

\begin{figure*}[]
\center
\includegraphics[width=0.49\textwidth]{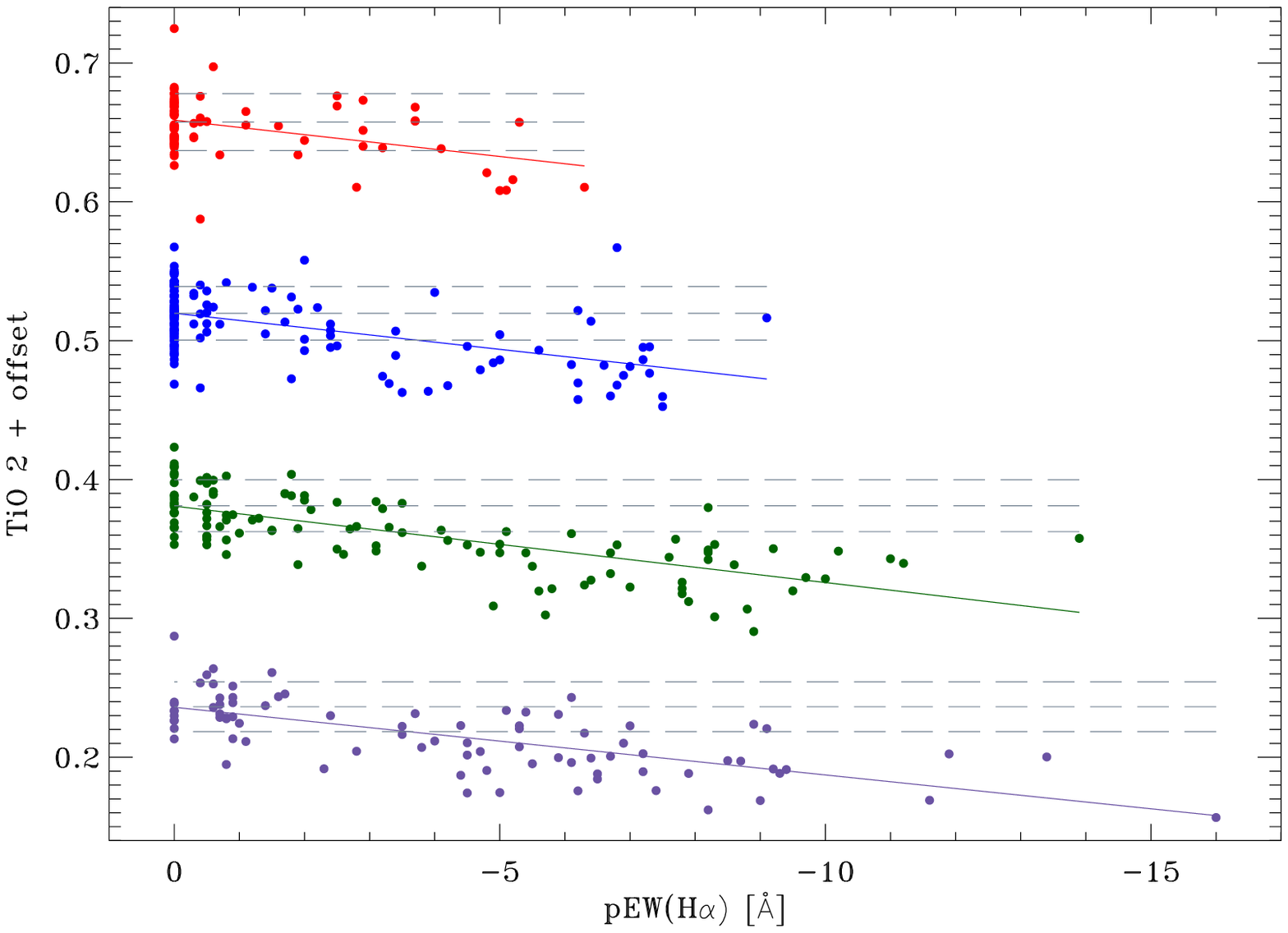}
\includegraphics[width=0.49\textwidth]{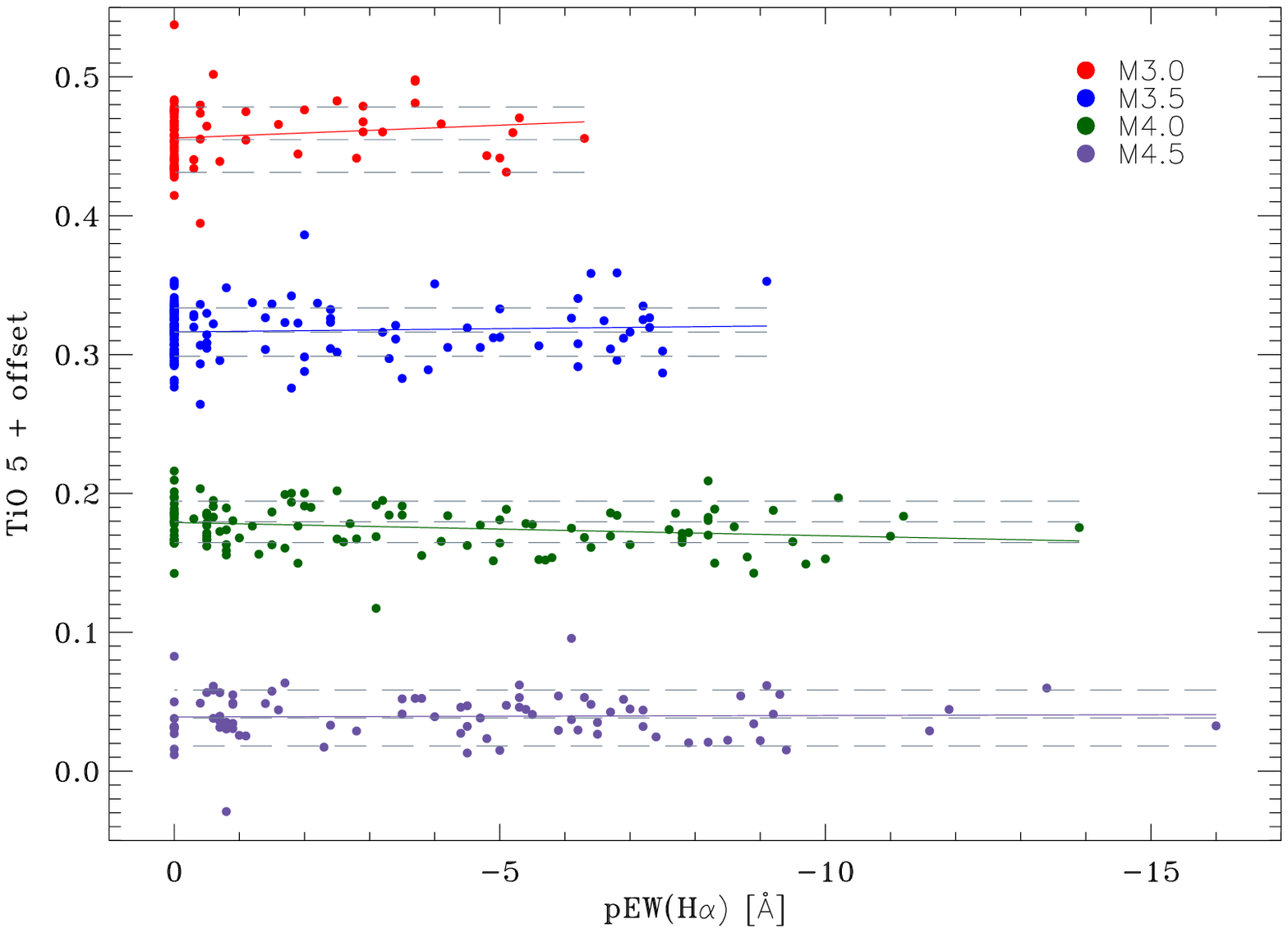}
\includegraphics[width=0.49\textwidth]{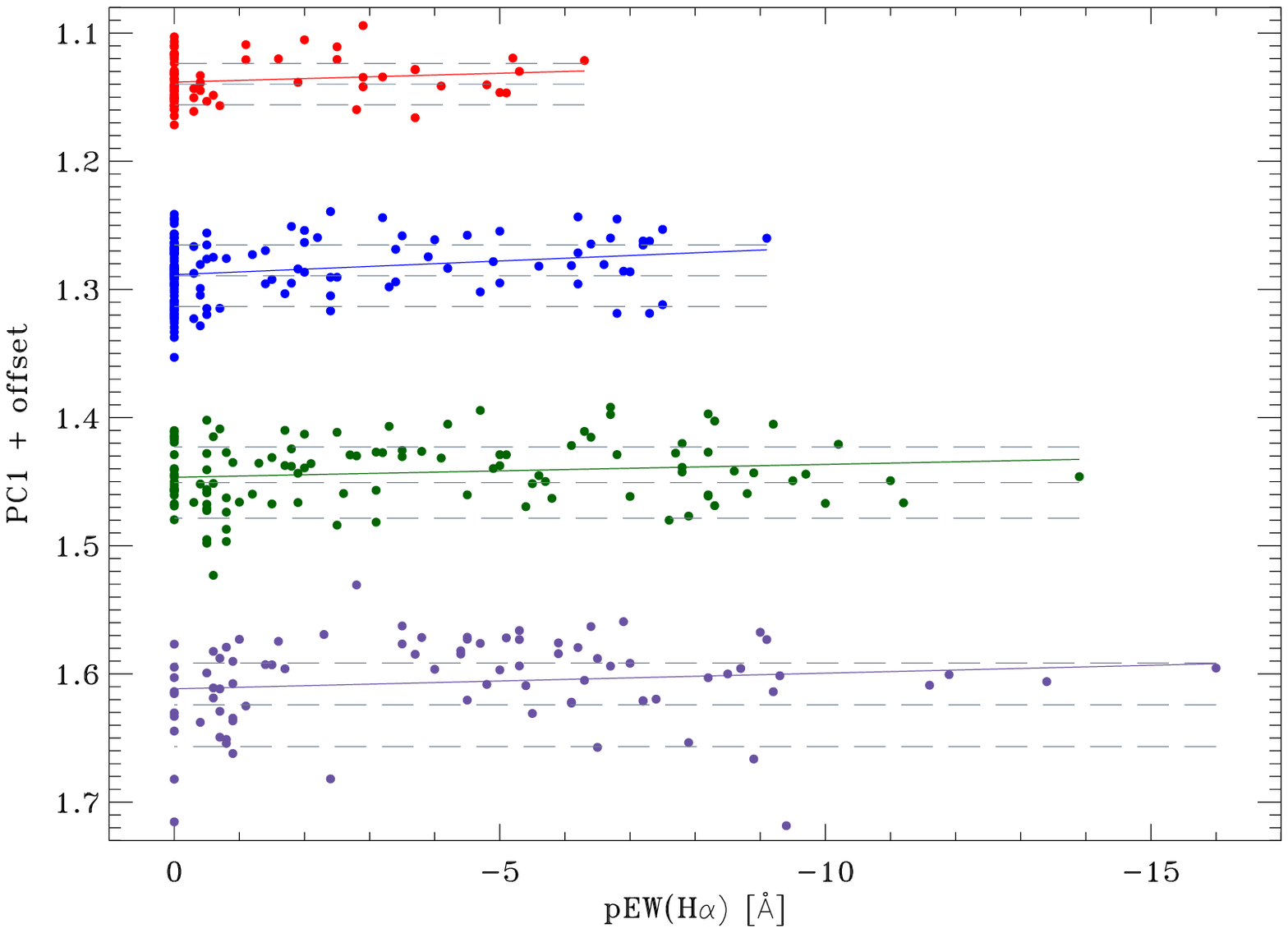}
\includegraphics[width=0.49\textwidth]{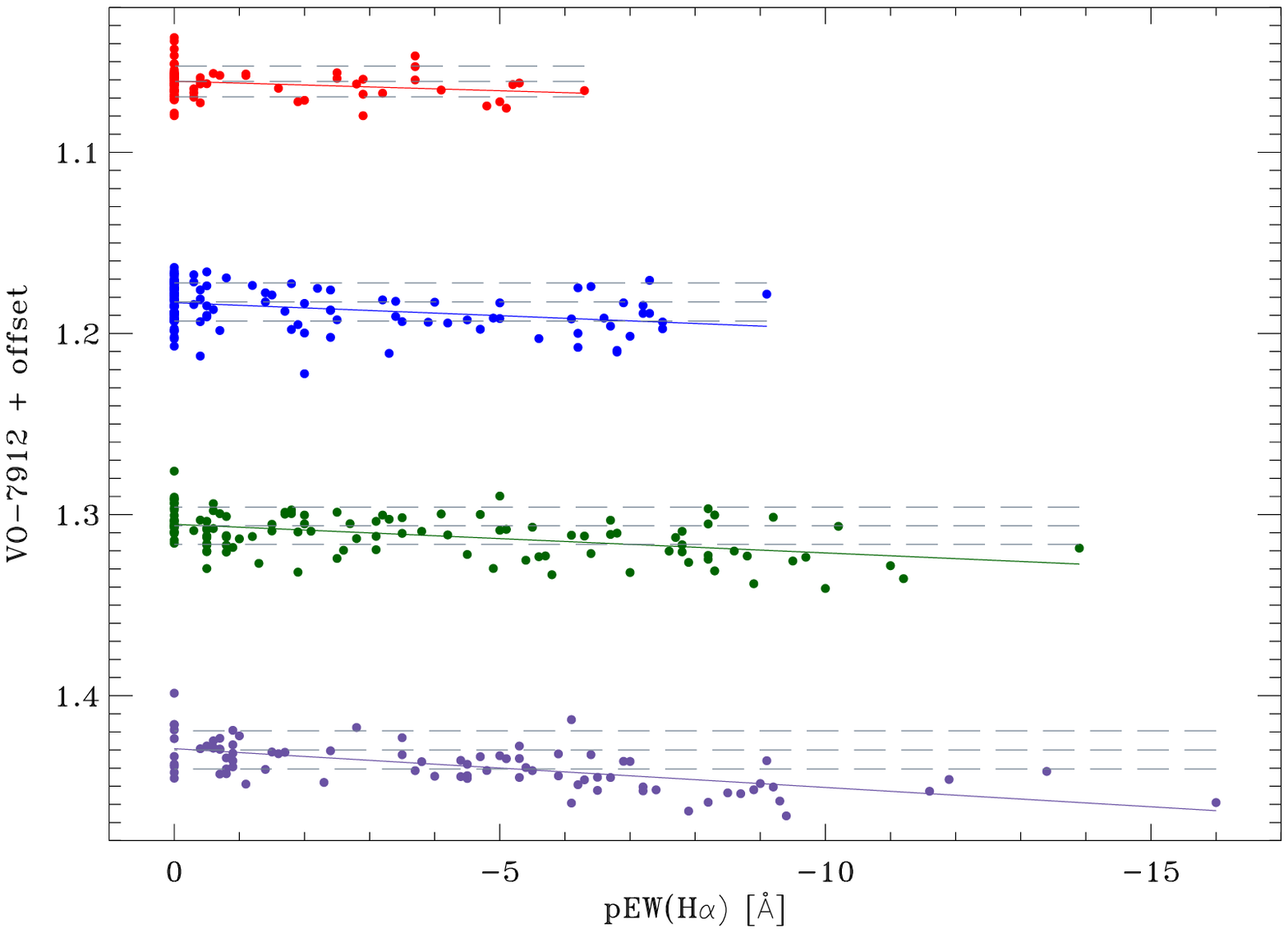}
\includegraphics[width=0.49\textwidth]{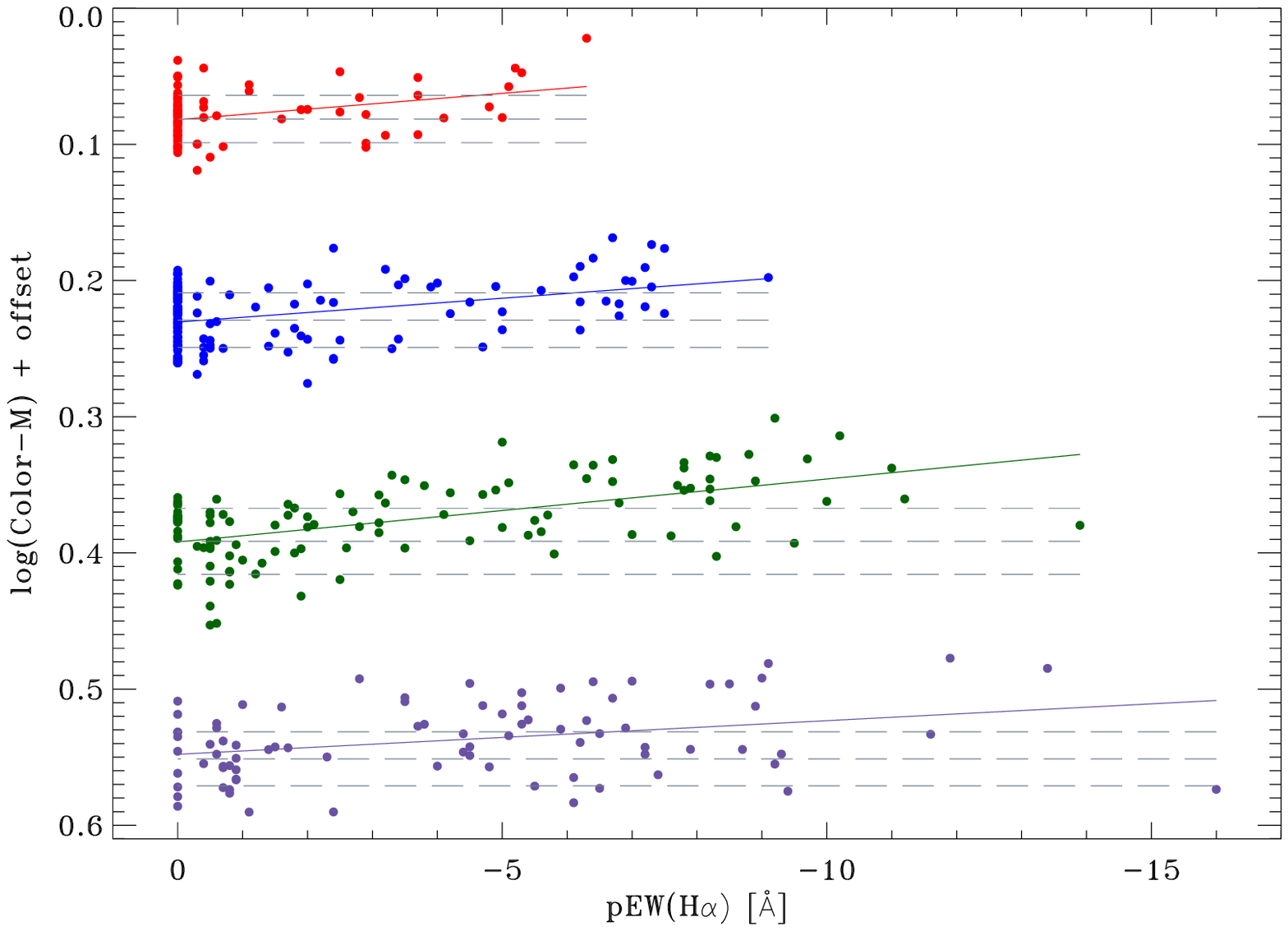}
\includegraphics[width=0.49\textwidth]{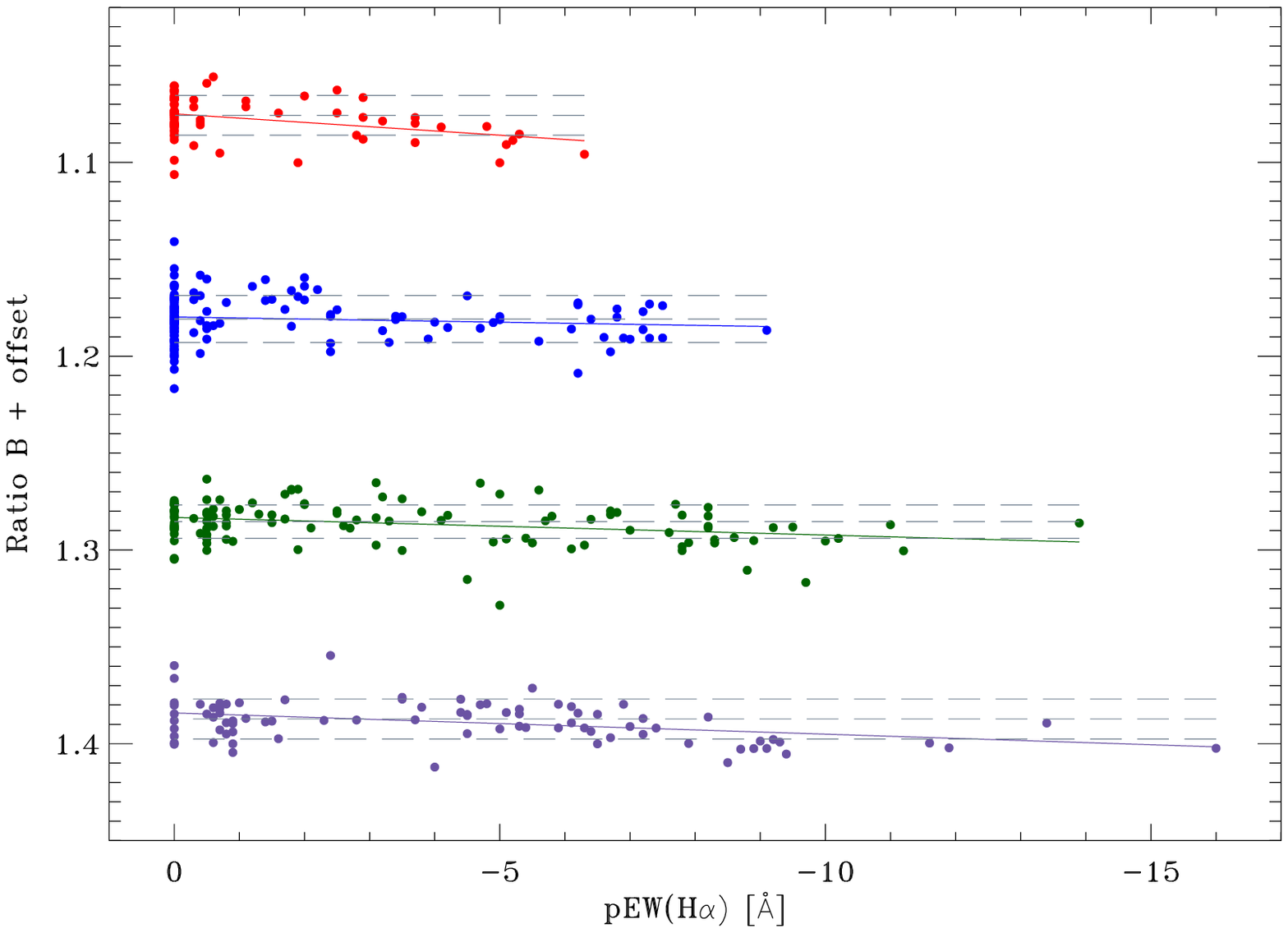}
\caption{\label{fig.IndicesvsHalpha} Spectral-typing indices TiO\,2, TiO\,5 ({\em top}), PC1, VO-7912 ({\em middle}), Color-M, and Ratio~B ({\em bottom}) as a function of H$\alpha$ pseudo-equivalent width for spectral types M3.0\,V, M3.5\,V, M4.0\,V, and M4.5\,V, from top to bottom.
For clarity, the indices are offset in steps of 0.1 in the vertical axis.
Solid lines are linear fits of the indices as a function of $pEW$H($\alpha$).
Dashed lines indicate the mean and $\pm 1 \sigma$ index at quiescence ($pEW$H($\alpha$) $>$ --1\,{\AA}).
} 
\end{figure*}

A large radius also translates into low gravity.
Indeed, the brightest of the trio of T~Tauri stars, J04313+241\,AB (V927~Tau\,AB, $J$ = 9.73\,mag) was the only non-giant target with spectral index Ratio~C $<$ 1.07 (Fig.~\ref{fig.RatioCvsPC1}) and the only one to which we did not assign a luminosity class in Table~\ref{table.SpT}.
Its optical spectrum is intermediate between those of giants and dwarfs of the same spectral type (M4.0:).  
Something similar is true for the other two T~Tauri stars, which also have very low Ratio~C indices for their spectral type (but all giant stars in our sample display H$\alpha$ in absorption).
Although T~Tauri stars are not natural targets for radial-velocity searches of low-mass exoplanets and none of the trio satisfies our criteria to be considered in the CARMENES sample (Table~\ref{table.carmencitalimits}), a {monitoring} of bright, young, M dwarfs could shed light on the process of exoplanet formation (e.g., Crockett et~al. 2012).
Furthermore, young, nearby, very late stars are also ideal targets for direct-imaging surveys for Jupiter-like planetary companions at wide separations (Masciadri et~al. 2005; Daemgen et~al. 2007; Chauvin et~al. 2010; Biller et~al. 2013; Delorme et~al. 2013, and references therein).
Some of these targets are 
J13143+733\,AB (NLTT~33370, M6.0\,V in AB~Doradus) and 
J09328+269 (DX\,Leo~B, M5.5\,V in Hercules-Lyra).

\subsubsection{Effect of activity on spectral typing}

As pointed out above, chromospheric activity could affect spectral typing. 
To study the validity of our results, we plot in Fig.~\ref{fig.IndicesvsHalpha} four of the five indices that we used for spectral typing as a function of $pEW$(H$\alpha$).
We investigated the four spectral type intervals with the largest number of stars (in parenthesis): M3.0\,V (72), M3.5\,V (134), M4.0\,V (113), and M4.5\,V (84). 
Grouping by spectral type minimised the natural variation of the spectral index with effective temperature.
The effect of activity on the TiO\,5 and PC1 indices is not significant.
However, strong activity in the largest quartile of $pEW$(H$\alpha$) has an appreciable effect on the indices VO-7912 and Color-M (not shown), but they fortuitously compensate each other because of the opposite slopes in their index vs. $pEW$(H$\alpha$) relations.
The effect of activity on the TiO\,2 index is more appreciable in the top left panel of Fig.~\ref{fig.IndicesvsHalpha}, {which agrees with the results shown by Hawley et~al. (1996), who  found that active M dwarfs tended to have lower values of TiO\,2 (more absorption) than inactive dwarfs with the same TiO\,5 indices (spectral types).
However, this level of activity} translates into a variation in spectral type of less than 0.8 subtypes for the most active stars, after using the coefficients in Table~\ref{table.fits}.
In the end, the counter-weighting combination of five spectral indices and, especially, the use of the $\chi^{2}$ and best-match methods guarantee that our adopted spectral types are free from the effect of chromospheric activity (for the investigated interval of $pEW$(H$\alpha$)).

Our results on quantifying the variation of some spectral types as a function of activity are seemingly in contrast to some previous works, such as Morales et~al. (2008).
However, a direct comparison should be avoided because they grouped the stars by absolute magnitude, for which a determination of the distance is needed.
A specific work on activity in M dwarfs will be another item of this series of papers on the science preparation of the CARMENES sample. 
It will be supported on one hand by new measurements of the emission of H$\alpha$, H$\beta$, and the Ca~{\sc ii} H \& K doublet and near-infrared triplet from high-resolution spectra and on the
other hand by a comprehensive parallax distance compilation and accurate spectro-photometric distance determination.

\section{Summary}

CARMENES, the new spectrograph at the 3.5\,m Calar Alto telescope, will spectroscopically monitor a sample of M dwarfs to detect exoplanets with the radial-velocity method. 
We are selecting the best planet host star candidates.
For that, we are compiling a comprehensive list of dwarf stars coming from existing spectroscopic and photometric catalogues, as well as from late-type star studies.
Currently, we  are gathering all available information and determine fundamental properties from observations for approximately 2200 targets.
Here we presented the first paper of a series that explains in detail the characterisation of our sample of targets for the CARMENES survey.
This paper detailed optical low-resolution spectroscopy.

One of the key stellar astrophysical parameters that we need for each target is its spectral type.
From the spectral type we estimate the stellar mass and infer planet detectability thresholds, and we ensure that we collate an even sampling of early-, mid-, and late-M dwarfs. 
Here, we undertook low-resolution spectroscopic observations of 753 targets with the CAFOS spectrograph on the 2.2\,m Calar Alto telescope.
This CAFOS sample contained M-dwarf candidates with poorly constrained spectral types, cool stars in multiple systems, and numerous comparison and standard stars.
We classified our targets using both least-squares fitting techniques and 31 spectral indices, of which we chose five indices with small dispersion to empirically calibrate spectral types (TiO\,2, TiO\,5, PC1, VO-7912 and Color-M).

Additionally, we investigated the relation of spectral indices with surface gravity.
We classified {25} of the observed targets as giant stars using the CaH series of related spectral indices, which are useful indicators to segregate giant stars from dwarfs. 
Metallicity was estimated through the $\zeta$ parameters (L\'epine et al. 2007). 
We concluded that all our field dwarf stars except two new subdwarf candidates have solar metallicity. 
We identified {49} late-type stars as young dwarfs in star-forming regions or moving groups already reported in the bibliography.

Finally, we also computed stellar activity indicators. 
Stellar activity is a fundamental property for the CARMENES survey because activity features, such as photospheric spots, can mimic the signature of exoplanets or increase the stellar intrinsic jitter that can mask real exoplanet signals. 
We computed the pseudo-equivalent width of the H$\alpha$ line of each target as an activity indicator, and analysed the effect of activity on spectral typing through indices. 
Although we have identified significant trends for some indices, the spectral type variation due to stellar activity is below one subtype level.

In summary, from the {753} targets that we observed with CAFOS, we obtained for the first time spectral types for {305} stars and improved it for {448} stars. 
We estimated gravity, metallicity, and activity indices for all targets. 
We identified {683} M dwarfs, of which {520} fulfill the CARMENES requirements and, therefore, will be included in the list of input targets. 
A more detailed {investigation} of these targets with high-resolution spectroscopic and imaging observations to select the best candidates for the CARMENES survey will produce the largest compilation of fully characterised M-type stars.

\begin{acknowledgements}

CARMENES is funded by the German Max-Planck-Gesellschaft (MPG), the Spanish Consejo Superior de Investigaciones Cient\'ificas (CSIC), the European Union through European Regional Fund (FEDER/ERF), the Spanish Ministry of Economy and Competitiveness, the state of Baden-W\"urttemberg, the German Science Foundation (DFG), and the Junta de Andaluc\'ia, with additional contributions by the members of the CARMENES Consortium (Max-Planck-Institut f\"ur Astronomie, Instituto de Astrof\'isica de Andaluc\'ia, Landessternwarte K\"onigstuhl, Institut de Ci\`encies de l'Espai, Institut f\"ur Astrophysik G\"ottingen, Universidad Complutense de Madrid, Th\"uringer Landessternwarte Tautenburg, Instituto de Astrof\'isica de Canarias, Hamburger Sternwarte, Centro de Astrobiolog\'ia, and the Centro Astron\'omico Hispano-Alem\'an).
Financial support was also provided by the {Universidad Complutense de Madrid}, the Comunidad Aut\'onoma de Madrid, the Spanish Ministerios de Ciencia e Innovaci\'on and of Econom\'ia y Competitividad, and the Fondo Europeo de Desarrollo Regional (FEDER/ERF) under grants 
AP2009-0187, 
SP2009/ESP-1496,        
AYA2011-30147-C03-01, -02, and -03, 
AYA2012-39612-C03-01, 
and ESP2013-48391-C4-1-R. 
Based on observations collected at the Centro Astron\'omico Hispano Alem\'an (CAHA) at Calar Alto, operated jointly by the Max--Planck Institut f\"ur Astronomie and the Instituto de Astrof\'{\i}sica de Andaluc\'{\i}a.
This research made use of the SIMBAD, operated at Centre de Donn\'ees astronomiques de Strasbourg, France, the NASA's Astrophysics Data System, the RECONS project database ({\tt http://www.recons.org}), the M, L, T, and Y dwarf compendium housed at {\tt http://dwarfarchives.org} maintained by C. Gelino, J. D. Kirkpatrick and A. Burgasser, and the Washington Double Star Catalog maintained at the U.S. Naval Observatory.
We thank S. L\'epine and E. Gaidos for sharing unpublished data with us, J.~I. Gonz\'alez-Hern\'andez, E.~W. Guenther, A. Hatzes, and M. R. Zapatero Osorio of the CARMENES Consortium for helpful comments, {and the anonymous referee for the quick and encouraging report}.

\end{acknowledgements}


\appendix

\section{Long tables}

\begin{list}{}{}
\item[$^{a}$] {\bf References to Table A.3} -- 
PMSU: Palomar/Michigan State University survey (see text);
Simbad: spectral type as reported by Simbad;
MP50: Moore \& Paddock 1950;    
Vys56: Vyssotsky 1956;                  
JM53: Johnson \& Morgan 1953;   
Lee84: Lee 1984;                                
Bid85: Bidelman 1985;                   
Ste86: Stephenson 1986;                 
SP88: Sanduleak \& Pesch 1988;  
Gar89: Garc\'ia 1989;                   
KMc89: Keenan \& McNeil et~al. 1989; 
Kir91: Kirkpatrick et~al. 1991;                 
Kri93: Krisciunas et~al. 1993;          
Hen94: Henry et~al. 1994;               
Jac94: Jacoby et~al. 1994;              
Mar94: Mart\'in et~al. 1994;            
Giz97: Gizis 1997;                              
GR97: Gizis \& Reid 1997;               
Mot97: Motch et~al. 1997;                       
App98: Appenzeller et~al. 1998; 
Gig98: Gigoyan et~al. 1998;             
Mot98: Motch et~al. 1998;                       
Cut00: Cutispoto et~al. 2000;           
Giz00: Gizis et~al. 2000a;                      
Li00: Li et~al. 2000;                           
CrRe02: Cruz \& Reid 2002;              
Gray03: Gray et~al. 2003;                       
Lep03: L\'epine et~al. 2003;            
Reid03: Reid et~al. 2003;               
Tee03: Teegarden et~al. 2003;           
Reid04: Reid et~al. 2004;               
Boc05: Bochanski et~al. 2005;           
Scho05: Scholz et~al. 2005;             
Gray06: Gray et~al. 2006;                       
Mon06: Montagnier et~al. 2006;  
Riaz06: Riaz et~al. 2006;                       
SB06: S\'anchez-Bl\'azquez et~al. 2006; 
Dae07: Daemgen et~al. 2007;             
Eis07: Eisenbeiss et~al. 2007;          
Reid07: Reid et~al. 2007;               
BS08: Bender \& Simon 2008;             
Jah08: Jahrei{\ss} et~al. 2008;         
Law08: Law et~al. 2008;                         
LC08: L\'opez-Corredoira et~al. 2008; 
Sce08: Scelsi et~al. 2008;              
Cab09: Caballero 2009;                  
Shk09: Shkolnik et~al. 2009;            
Cab10: Caballero et~al. 2010;           
Shk10: Shkolnik et~al. 2010;            
LG11: L\'epine \& Gaidos 2011;  
Jan12: Janson et~al. 2012;              
JE12: Jim\'enez-Esteban et~al. 2012; 
RA12: Rojas-Ayala et~al. 2012;  
Fri13: Frith et~al. 2013;                       
Lep13: L\'epine et~al. 2013;            
Mann13: Mann et~al. 2013;               
Abe14: Aberasturi et~al. 2014;          
Lam14: Lamert 2014;                     
New14: Newton et~al. 2014;              
RS14: Reyes-S\'anchez 2014.             
\end{list}

\scriptsize
\begin{longtable}{l  l  l  l c c c c c}
\label{table.observed}\\
\caption[]{Observed stars: identification, common name, Gliese number, 2MASS coordinates and $J$ magnitude, observing date, and exposure time.}\\
   \hline
   \hline
   \noalign{\smallskip}
No.	& Karmn			& Name	        				& Gl/GJ		& $\alpha$ 		& $\delta$			& $J$	&  Observation		&  $N \times t_{\rm exp}$ 	\\
	&     				& 						& 			& (J2000)			& (J2000)			& [mag]	&  date			&  [s]		        			\\
\noalign{\smallskip}
    \hline
    \noalign{\smallskip}		
 \endfirsthead
\caption[]{Observed stars: identification, common name, Gliese number, 2MASS coordinates and $J$ magnitude, observing date, and exposure time (cont.).}\\ 
  \hline
  \hline
  \noalign{\smallskip}		
No.	& Karmn			& Name	        				& Gl/GJ		& $\alpha$ 		& $\delta$			& $J$	&  Observation		&  $N \times t_{\rm exp}$ 	\\
	&     				& 						& 			& (J2000)			& (J2000)			& [mag]	&  date			&  [s]		        			\\
 \noalign{\smallskip}
  \hline
  \noalign{\smallskip}
  \endhead
  \noalign{\smallskip}
  \hline
  \endfoot
1 	&  J00066--070\,AB	&  2MASS J00063925--0705354 &  ... 		&  00:06:39.20		&  --07:05:35.3		& 9.83 	&   04 Aug 2012	&   1\,$\times$\,1000	\\ 
2 	&  J00077+603\,AB  	&  G 217--032	       			&  ... 			&  00:07:42.60    	&  +60:22:54.3    	& 8.91 	&   24 Sep 2012  	&   1\,$\times$\,600 \\ 
3 	&  J00115+591      	&  LSR J0011+5908         		&  ... 			&  00:11:31.82    	&  +59:08:40.0    	& 9.95 	&   11 Jan 2012  	&   2\,$\times$\,700 \\ 
4 	&  J00118+229      	&  LP 348--40	       			&  ... 			&  00:11:53.03    	&  +22:59:04.7    	& 8.86 	&   07 Dec 2011  	&   1\,$\times$\,250 \\  		      
5 	&  J00119+330      	&  G 130--053	       			&  ... 			&  00:11:56.54    	&  +33:03:17.8    	& 9.07	&   07 Dec 2011  	&   1\,$\times$\,220 \\
6 	&  J00122+304      	&  1RXS J001213.6+302906    	&  ... 			&  00:12:13.41    	&  +30:28:44.3    	& 10.24 	&   12 Nov 2011  	&   1\,$\times$\,600 \\
7 	&  J00133+275      	&  [ACM2014] J0013+2733 	&  ... 			&  00:13:19.52 		&  +27:33:31.1   	& 10.43 	&   12 Nov 2011  	&   1\,$\times$\,900 \\ 
8 	&  J00136+806      	&  G 242--048	       			&  3014 A		&  00:13:38.71    	&  +80:39:56.8    	& 7.76 	&   01 Sep 2012  	&   1\,$\times$\,300 \\
\end{longtable}


\scriptsize
\begin{longtable}{l ccccc cc c c}
\label{table.indices}\\
\caption[]{Seven representative spectral indices, $\zeta$ metallicity index, and H$\alpha$ pseudo-equivalent width.} \\
   \hline
   \hline
   \noalign{\smallskip}
Karmn	&  PC1	& TiO\,2	& TiO\,5	& VO-7912	& Color-M		& CaH\,2	& CaH\,3	& $\zeta$	& $pEW$(H$\alpha$) [\AA]	\\	 	
 \noalign{\smallskip}
    \hline
    \noalign{\smallskip}		
 \endfirsthead
  \caption[]{Seven representative spectral indices, $\zeta$ metallicity index, and H$\alpha$ pseudo-equivalent width (cont.).}\\ 
  \hline
  \hline
  \noalign{\smallskip}		
Karmn	&  PC1	& TiO\,2	& TiO\,5	& VO-7912	& Color-M		& CaH\,2	& CaH\,3	& $\zeta$	& $pEW$(H$\alpha$) [\AA]	\\	 	
 \noalign{\smallskip}
  \hline
  \noalign{\smallskip}
  \endhead
  \noalign{\smallskip}
  \hline
  \endfoot
J00066--070\,AB   &   1.269   &   0.492   &   0.317   &   1.148   &   1.778  &   0.404   &   0.654  &   0.973  &   $-2.3_{-0.5}^{+0.3}$    \\
\noalign{\smallskip}
J00077+603\,AB   &   1.198   &   0.532   &   0.369   &   1.111   &   1.405  &   0.408   &	0.631  &   0.883  &   $-6.7_{-0.4}^{+0.3}$    \\
\noalign{\smallskip}
J00115+591   &   1.511   &   0.378   &   0.202   &   1.222   &   2.902  &   0.281   &	0.564  &   0.970  &   $-1.6_{-0.4}^{+0.2}$    \\
\noalign{\smallskip}
J00118+229   &   1.215   &   0.606   &   0.405   &   1.090   &   1.404  &   0.492   &	0.751  &   1.052  &   $-0.5_{-0.2}^{+0.2}$    \\
\noalign{\smallskip}
J00119+330   &   1.167   &   0.634   &   0.427   &   1.072   &   1.293  &   0.503   &	0.748  &   1.023  &   $-0.3_{-0.2}^{+0.1}$    \\
\noalign{\smallskip}
J00122+304   &   1.296   &   0.497   &   0.354   &   1.154   &   1.755  &   0.427   &	0.685  &   0.972  &   $-8.7_{-0.5}^{+0.4}$    \\
\noalign{\smallskip}
J00133+275   &   1.296   &   0.512   &   0.339   &   1.144   &   1.805  &   0.425   &	0.686  &   0.994  &   $-4.0_{-0.4}^{+0.2}$    \\
\noalign{\smallskip}
J00136+806   &   1.027   &   0.770   &   0.601   &   1.012   &   0.901  &   0.644   &	0.812  &   1.002  &   $+0.0_{-0.2}^{+0.2}$    \\
\end{longtable}


\scriptsize
\begin{longtable}{lll cc ccccc l}
\label{table.SpT}\\
\caption[Spectral types of observed stars]{Spectral types of observed stars.}\\
   \hline
   \hline
   \noalign{\smallskip}
Karmn	& Sp. type		& Ref.$^{a}$	&  \multicolumn{2}{c}{Sp. type}	& \multicolumn{5}{c}{Sp. type}				& Sp. type \\
\cline{4-5}
\cline{6-10}
\noalign{\smallskip}
		& biblio. 		&           	& Best-fit		& $\chi^{2}$      	& TiO2 & TiO5 & PC1 & VO-7912 & Color-M	& adopted \\
 \noalign{\smallskip}
    \hline
    \noalign{\smallskip}		
 \endfirsthead
  \caption[]{Spectral types of observed stars (cont.).}\\ 
  \hline
  \hline
  \noalign{\smallskip}		
Karmn	& Sp. type		& Ref.$^{a}$	&  \multicolumn{2}{c}{Sp. type}	& \multicolumn{5}{c}{Sp. type}				& Sp. type \\
\cline{4-5}
\cline{6-10}
\noalign{\smallskip}
		& biblio. 		&           	& Best-fit		& $\chi^{2}$      & TiO2 & TiO5 & PC1 & VO-7912 & Color-M	& adopted \\
 \noalign{\smallskip}
  \hline
  \noalign{\smallskip}
  \endhead
  \noalign{\smallskip}
  \hline
\endfoot
 \noalign{\smallskip}
  \hline
\endlastfoot
J00066--070\,AB & M3.5\,V+m4.5: & Reid07, Jan12 & 4.5 & 4.5 & 4.5 & 4.5 & 4.0 & 4.5 & 4.5 &  M4.5\,V \\ 
J00077+603\,AB & M4.5\,V & Lep13 & 4.0 & 4.0 & 4.5 & 4.0 & 3.5 & 4.0 & 3.5 &  M4.0\,V \\
J00115+591 & M5.5\,V & Lep03 & 6.0 & 6.0 & 5.5 & 6.0 & 5.5 & 5.5 & 5.5 &  M5.5\,V \\
J00118+229 & M3.5\,V & Reid04 & 3.5 & 3.5 & 3.5 & 3.5 & 4.0 & 4.0 & 3.5 &  M3.5\,V \\
J00119+330 & M3.5\,V & Giz97 & 3.5 & 3.0 & 3.5 & 3.5 & 3.5 & 3.5 & 3.5 &  M3.5\,V \\
J00122+304 & M5.0\,V & Abe14 & 4.5 & 4.5 & 4.5 & 4.0 & 4.5 & 5.0 & 4.5 &  M4.5\,V \\
J00133+275 & M4.5\,V & Abe14 & 4.5 & 4.5 & 4.5 & 4.5 & 4.5 & 4.5 & 4.5 &  M4.5\,V \\
J00136+806 & M1.5\,V & PMSU & 1.5 & 1.5 & 1.5 & 1.5 & 1.5 & 1.5 & 1.5 &  M1.5\,V \\
\end{longtable}
%
\normalsize

\end{document}